\documentclass[12pt,preprint]{aastex}
\usepackage{mkfig}

\begin{document}

\title{\bf Imaging Prominence Eruptions Out to 1 AU}

\author{Brian E.\ Wood, Russell A.\ Howard, \& Mark G.\ Linton}

\affil{Naval Research Laboratory, Space Science Division,
  Washington, DC 20375}
\email{brian.wood@nrl.navy.mil}


\begin{abstract}

     Views of two bright prominence eruptions trackable all the
way to 1~AU are here presented, using the heliospheric imagers on the
{\em Solar TErrestrial RElations Observatory} (STEREO) spacecraft.
The two events first erupted from the Sun on 2011~June~7 and
2012~August~31, respectively.  Only these two examples of
clear prominence eruptions observable this far from the Sun could
be found in the STEREO image database, emphasizing the rarity
of prominence eruptions this persistently bright.  For the 2011~June
event, a time-dependent 3-D reconstruction of the prominence structure
is made using point-by-point triangulation.  This is not possible for
the August event due to a poor viewing geometry.
Unlike the coronal mass ejection (CME) that accompanies it,
the 2011~June prominence exhibits little deceleration from the Sun to
1~AU, as a consequence moving upwards within the CME.  This
demonstrates that prominences are not necessarily tied to the CME's
magnetic structure far from the Sun.  A mathematical framework
is developed for describing the degree of self-similarity for
the prominence's expansion away from the Sun.  This analysis
suggests only modest deviations from self-similar expansion, but
close to the Sun the prominence expands radially somewhat more
rapidly than self-similarity would predict.

\end{abstract}


\section{INTRODUCTION}

     The heliospheric imagers on the two {\em Solar TErrestrial RElations
Observatory} (STEREO) spacecraft began monitoring the Sun-Earth line
shortly after the launch of STEREO in 2006~October.  Since then, they
have demonstrated their ability to track coronal mass ejections (CMEs)
from the Sun all the way to 1~AU \citep{raha08,bew09b,cjd09,nl12,cm14}.
In addition to tracking CMEs, they have also provided the
first images of corotating interaction regions (CIRs) moving through
the inner heliosphere \citep{nrs08,apr08,bew10b}.
More rarely, prominence eruptions can also be
observed in images out to 1~AU \citep{tah15b}, and this phenomenon is
here studied in more detail.

     Prominence eruptions are often accompanied by CMEs.
Although CMEs come in a wide variety of sizes and shapes
\citep{rah85}, the most structured and organized of
them often have a three-part structure in white light coronagraph
images, consisting of a bright circular rim, a dark cavity within the
rim, and a bright core typically near the back of the cavity.  This
bright core is often interpreted as being prominence material, but
CMEs with this structure are observed without accompanying prominence
eruptions, so the nature of the core can be ambiguous.  There
are cases of erupting prominences that can be clearly tracked upwards
into the fields of view of coronagraphs \citep[e.g.,][]{ns00},
in which the prominence material tends to be bright and
concentrated.  However, even when such cases are observed, the
prominence emission often fades and becomes difficult to distinguish
from the rest of the CME structure by the time it leaves the
coronagraph field of view.  The two events presented in this article,
from 2011~June~7 and 2012~August~31, are particularly bright events
which maintain their recognizable appearance in white light images
from close to the Sun all the way to 1~AU.

\section{OBSERVATIONS OF THE PROMINENCE ERUPTIONS}

     Prominence eruptions are most naturally studied near the Sun
in ground-based H$\alpha$ images, or from space in the EUV.  Currently,
the most useful observing platform for such studies
is the Atmospheric Imaging Assembly (AIA) instrument on board the
{\em Solar Dynamics Observatory} (SDO).  Figure~1 shows SDO/AIA images of
the two prominence eruptions of interest here, from 2011~June~7 and
2012~August~31.  The former was accompanied by an M2 class X-ray flare
and the latter by a C8 flare.  The 2011 June event is a very well
studied event, being especially notable for the large amount of dark
prominence material that rains back down onto the surface after the
eruption \citep{dei12,drw13,fr13,hrg13,jc14}.  The 2012
August event has not garnered as much attention, but \citet{tah15a,tah15b}
used it to study the relative importance of H$\alpha$ emission and
Thomson scattering in white-light coronagraphic images of the erupting
prominence and, unlike the studies of the 2011~June event, also
followed its evolution into the inner heliosphere.  Unlike the messy
2011 June eruption, the 2012 August event involves the ejection of a
single, large, well-defined, loop-like filament that had been visible
for weeks before finally erupting on 2012 August 31.  Both of the two
prominence eruptions are accompanied by bright, fast CMEs, which are
trackable all the way to 1~AU along with the prominence material.  The
CME structures and their kinematics will be studied here along with
the prominence eruptions.

     This article mostly concerns observations of the prominence
eruptions (and CMEs) far from the Sun, particularly with the white-light
telescopes on STEREO.  Each STEREO spacecraft carries four such
telescopes that observe at different distances from the Sun \citep{raho08}.
There are two coronagraphs, COR1 and COR2, that observe
the Sun's white light corona at angular distances from Sun-center of
$0.37^{\circ}-1.07^{\circ}$ and $0.7^{\circ}-4.2^{\circ}$, respectively,
corresponding to distances in the plane of the sky of
$1.4-4.0$ R$_{\odot}$ for COR1 and $2.5-15.6$~R$_{\odot}$ for COR2.
And there are two heliospheric imagers, HI1 and HI2 \citep{cje09},
that monitor the interplanetary medium (IPM) in between the Sun and
Earth, where HI1 observes elongation angles from Sun-center of
$3.9^{\circ}-24.1^{\circ}$ and HI2 observes from $19^{\circ}-89^{\circ}$.

     The two STEREO spacecraft have orbits near 1~AU, with STEREO-A
leading the Earth by a steadily increasing angular distance, and
STEREO-B trailing behind.  Figure~2 maps the locations of Earth,
STEREO-A, and STEREO-B in the ecliptic plane at the time of the two
prominence eruptions.  The fields of view of the HI1 and HI2 imagers
are shown explicitly in the figure.  Outlines of the CME structures
are from 3-D reconstructions that will be described below.

     Figure~3 displays sequences of images illustrating how STEREO
can follow the two prominence eruptions far into the IPM.  The COR1
and COR2 images are displayed after subtracting an average daily
background.  The HI1 and HI2 images are running-difference images,
with the previous image subtracted from each image.  For the 2011 June
event, STEREO-A is best situated to track the eruption (see Figure~2),
so a sequence of eight STEREO-A images is shown.  For the 2012 August
event, STEREO-B is the better situated platform, so eight STEREO-B
images are shown.  Movies of the prominence eruptions, cycling through
all four of STEREO's white-light imagers, are available in the online
version of this article.

     The two prominence eruptions are so bright that they actually
saturate the COR1 detector, as shown in the first COR1 images
of each event in Figure~3.  This COR1 saturation is not common at all,
so it is clear early on that these eruptions are exceptional in how
bright they appear in the upper corona.  The COR2 and HI1
images in Figure~3 help to place the prominence eruptions in the
context of the larger CME eruptions in which they are embedded.

     Colored arrows in Figure~3 are used to point to specific
parts of the prominences that can be tracked from frame to frame.
The most recognizable part of the 2011 June eruption is a
square-topped loop.  A red arrow points to the top of the southern
leg of this loop.  The 2012 August eruption is initially a more
circular loop, as seen most clearly in the COR2-B images in
Figure~3.  However, the top of this loop and its northern leg
quickly fade and become difficult to distinguish from the CME
background structure.  This is typical for prominence eruptions
in white light.  It is the bright southern leg of the loop that
is easily trackable from COR1-B through HI2-B, and this is what
the green arrow in Figure~3 is following.

     The blue arrows in Figure~3 identify 2011 June prominence
material ejected ahead of the aforementioned square-topped loop.
This material is trackable at least into HI1 before becoming more
diffuse and less recognizable as distinct from the background
CME.  Once again, this is the norm for prominence material.
What is remarkable about the two prominence eruptions presented
here are the parts of the prominence structures that remain
bright, distinct, compact, and coherent all the way out to 1~AU.

     The yellow arrows in Figure~3 point to two dense knots of
material far down the northern leg of the square-topped loop of the
2011 June eruption.  These knots actually begin to look like comets as
they travel through the HI1-A field of view, with tails pointed
away from the Sun.  This is most easily seen
in the movie of the 2011 June prominence eruption available in the
online version of this article, but the tails of the ``comets'' are
faintly visible in the third HI1-A image in Figure~3.  This phenomenon
indicates that these knots of material are moving more slowly than the
ambient solar wind at this location.  The knots are clearly dense
enough that the solar wind is having difficulty picking them up and
accelerating them to its speed; hence the development of the ``comet
tails.''

     It would be nice if the extensive observations of these
prominence eruptions from STEREO's imagers could be supplemented with
direct plasma measurements from in~situ instruments on STEREO, or from
the {\em Wind} and Advanced Composition Explorer (ACE) spacecraft
operating near Earth.  The images clearly show the prominence material
persisting to great distances from the Sun, so the in~situ signal
from such an encounter should have been quite obvious and dramatic.
Unfortunately,
the prominence eruptions missed all of the 1~AU spacecraft, with the
2011~June~7 eruption passing in between Earth and STEREO-A and the
2012~August~31 eruption passing between Earth and STEREO-B (see
Figure~2).

\section{CME RECONSTRUCTION}

     Given that the erupting prominences are incorporated into larger
CME structures, a 3-D reconstruction of the two CMEs is useful
to quantify the trajectory, spatial extent, and kinematics of
the larger scale eruptions.  Representative images of the CMEs are shown
in Figure~4.  These include not only images from STEREO-A and -B,
but also from the C2 coronagraph on the Large Angle Spectrometric
Coronagraph (LASCO) instrument on board the {\em Solar and Heliospheric
Observatory} (SOHO), which operates near Earth \citep{geb95}.
Observations from all three spacecraft are considered in the
reconstruction analysis.

     The 3-D reconstructions are performed using techniques
described and used extensively in past work \citep{bew09a,bew10a,bew11},
and the reader is referred to those papers for
details.  In short, the CME ejecta is assumed to have the shape of a
symmetric magnetic flux rope \citep[e.g.,][]{jc97,seg98,bew99,at09}.
Both of these fast CMEs have visible shock
fronts out in front of the ejecta, so the 3-D shock shape is
reconstructed as well, approximating it as a symmetric, lobular front.
The CME structure is not assumed to vary at all with
time, meaning we assume the structure expands in a self-similar
fashion.  There is clearly some change in structure in reality, so the
trial-and-error fitting process involves compromise in deciding on a
best estimate of the average CME shape from close to the Sun (in COR1
and COR2) into the IPM (in HI1 and HI2).  The resulting CME structures
are shown in Figure~5, each structure once again
consisting of a tube-shaped magnetic flux rope embedded inside a
lobular shock front.  The structures are shown in heliocentric
Earth ecliptic (HEE) coordinates, with the Sun at the origin, the
x-axis pointed towards Earth, and the z-axis pointed towards ecliptic
north.  Slices through these 3-D models in the ecliptic
plane are shown in Figure~2, providing another good visualization of
the trajectory and spatial extent of these two CMEs.  Synthetic images
of the reconstructions are shown in Figure~4 for comparison with the
actual images.

     The reconstruction process includes a kinematic analysis of
the CMEs, once again completely analogous to what has been done in
past work \citep{bew09a,bew10a,bew11}.  For both events,
the leading edge of the flux rope component of the CME is
followed from the Sun to beyond 1~AU, with angular distance
from the Sun converted to actual distance using the so-called
``harmonic mean'' approximation \citep{nl09}.  The resulting
distances are plotted versus time in the upper panels of Figure~6,
for both CMEs.  These distance-vs.-time measurements are fitted
with a simple, multi-phase kinematic model.  We have found that
fast, impulsive CMEs generally have kinematic profiles that can
be modeled with three phases:  an constant acceleration phase very
close to the Sun; a constant deceleration phase as the CME
decelerates, presumably due to interaction with the slower ambient
solar wind; and finally a constant velocity phase after the
CME reaches its terminal velocity \citep{bew09a,bew11,bew12}.
We model our two CMEs similarly, but the 2011~June~7 CME is
already at its maximum velocity early in COR1, so the first phase
is omitted.  The result is the
fit to the data in the top panels of Figure~6, and the
velocity profiles in the bottom panels.  Both CMEs are quite fast,
reaching speeds of 1012 km~s$^{-1}$ and 1244 km~s$^{-1}$,
respectively.  These maximum velocities are achieved very early,
even before they leave the COR1 field of view.  Both CMEs
then decelerate significantly, mostly in the HI1 field of view.

     The reconstructions suggest that the 2011~June~7 CME grazes
Earth, and that the 2012~August~31 CME grazes both Earth and
STEREO-B (see Figure~2).  Figure~7 shows in~situ plasma measurements
from {\em Wind} (near Earth) during the presumed 2011~June~7 CME
encounter, and from both {\em Wind} and STEREO-B during the
2012~August~31 CME encounters.  Both the proton density and
velocity measurements are shown, which are compared with
the predicted density and velocity profiles from the 3-D
reconstruction and kinematic models of the two CMEs.

     For the 2012~August event, clear shocks are observed at both
{\em Wind} and STEREO-B on September 3 at times when the
model CME is predicted to encounter the spacecraft, providing
strong support for these shocks being the 2012~August~31 event.
The velocity predictions appear much too high, but comparing
image and in~situ velocity measurements for shocks can be tricky,
because the images will be measuring the velocity of the shock,
while the in~situ data will be measuring the velocity of material
moving through the shock.  The agreement is therefore deemed good
enough for our purposes.
The negative slope of the predicted velocity
profiles in Figure~7 is due to the self-similar expansion
assumption in the 3-D reconstruction process.
Evidence that the 2011~June~7 CME encounters {\em Wind}
is much weaker.  There is a modest density enhancement on
June~10, which persists for about a day, but there is no shock
or velocity increase at the beginning of the density increase.
The density increase begins a few hours after the reconstruction
predicts the CME shock should arrive, which is close enough to
suspect that this density increase is associated with the CME.

    For the 2012~August event, the existence of clear shocks at
both STEREO~B and {\em Wind} provides another excellent way to
confront our image-based reconstruction with in~situ data.  By
combining the in~situ measurements and the Rankine-Hugoniot
shock jump conditions, shock normals at STEREO-B and {\em Wind} can
be inferred, which can then be compared with shock normals predicted
by the shock shape in Figure~5.  This is actually an ideal event
for this kind of exercise, because STEREO-B and {\em Wind} are on
opposite sides of the shock and should therefore see shock
normals pointing in very different directions (see Figure~2).

     The shock normal computation procedure is described in
\citet{bew12}, based largely on the analysis approach of
\citet{ak08}, who describe the
methodology in even more detail.  Figure~8 shows
proton density, temperature, velocity, and magnetic field
measurements from STEREO-B and {\em Wind} taken at the time
of the shock encounter on 2012~September~3.  Pre- and post-shock
time periods are delineated and used to measure pre- and post-shock
plasma parameters, with uncertainties estimated using the variation
of the parameters within the defined 5--10 minute time range.

     These pre- and post-shock plasma measurements are plugged
into the Rankine-Hugoniot shock jump
conditions \citep[e.g.,][]{rpl71}.  Different
shock normals can be defined by a longitude $\phi$ and latitude
$\theta$ in a Sun-centered coodinate system, with
the spacecraft at $(\phi,\theta)=(0^{\circ},0^{\circ})$, $\theta=90^{\circ}$
being to the north relative to the ecliptic, and $\phi=90^{\circ}$
being to the west (i.e., to the right from the spacecraft's
perspective).  The goal is to determine which shock
normal best meets the jump conditions, as quantified by a
$\chi^2$ goodness-of-fit parameter.  As described by
\citet{ak08}, the shock velocity normal to the
shock in the rest frame of the shock, $V_S$, is generally another
free parameter of the fit.  However, this parameter can be
eliminated using kinematic information from imaging \citep{bew12}.
The 3-D shock reconstruction
and kinematic model described above provide a measure of shock
radial velocity at the spacecraft, $V_{rad}$.
From Figure~7, $V_{rad}=681$~km~s$^{-1}$ at STEREO-B
and $V_{rad}=589$~km~s$^{-1}$ at {\em Wind}.  The radial velocity is
related to $V_S$ by $V_S=V_{rad}\cos\phi \cos\theta$.

     Figure~9 shows $\chi^2$ contours of the shock normal fits
in the $\theta-\phi$ parameter space, and compares these measurements
with the shock normals expected from the 3-D reconstruction.
As expected, the shock normal is pointed to the east at STEREO-B,
which is near the east edge of the shock, and the shock normal
is pointed to the west at {\em Wind}, which is near the west edge
of the shock (see Figure~2).  The jump condition measurements of
the shock normals agree very well with those predicted by the
image-based shock reconstruction, with only a $21^{\circ}$ discrepancy
at STEREO-B and a $33^{\circ}$ discrepancy at {\em Wind}.  This
result provides support for the accuracy of our 3-D CME reconstruction,
and demonstrates admirable consistency between the imaging and
in~situ measurements of the CME.

\section{PROMINENCE RECONSTRUCTION}

     The CME reconstruction process described in the
previous section has been used in the reconstruction of a number
of past events, but it does involve a number of assumptions and
approximations.  The analysis assumes that the CME structure is
the same at all times, that it can be perfectly represented by the
simple symmetric flux-rope-plus-shock paradigm, and that expansion
is purely radial and self-similar.  None of these assumptions
are strictly true, as CME structures always have asymmetries and
irregularities to their shapes that will be impossible for any
parametrized representation to accurately reproduce, and CMEs
often exhibit some degree of shape variation during their journey
from the Sun to 1 AU.
For the prominence, we utilize a 3-D reconstruction approach
that does not require any a priori assumptions about the shape or
time-dependence of the structure.

      Specifically, the analysis approach is a point-by-point
reconstruction of the time-dependent erupting prominence structure
using simple triangulation from two-viewpoint STEREO images.  The
basic idea behind the use of two-viewpoint STEREO imagery for
triangulation in 3-D reconstructions is described by \citet{pcl09},
who focused on the 3-D reconstruction of an erupting filament close to
the Sun.  The technique requires simultaneous images, image A and
image B, of some structure taken from two different vantage points.  A
point in image A marking the location of some feature corresponds to a
line in 3-D space, with the feature lying somewhere along that line.
The line in 3-D can be mapped onto a line in the image plane of image
B.  The game is then to identify where along the image B line is the
structure marked in image A.  Once this is known, the 3-D coordinates
of the feature are known.  \citet{pcl09} use the word ``tiepointing''
to describe this sort of analysis.  Doing this for multiple points
allows a crude mapping of the 3-D structure that is being observed.

     An example is shown in Figure~10, which displays COR2-A and
COR2-B images of the 2011~June event, taken at the same
time.  Fifteen points on the prominence structure are identified,
indicating the features to be tracked as the prominence expands
away from the Sun.  The first ten points trace the square-topped
loop that defines the most recognizable part of the structure.
Point \#5 basically corresponds to the red arrow in Figure~3, and
points \#9 and \#10 correspond to the yellow arrows.
Points \#11-\#15 identify other features that are not part of this
loop, which triangulation reveals are in front of the loop from
STEREO-A's perspective.  The connectivity of these last five
points with each other and with the loop is unclear.  In any case,
point \#5 in the COR2-A image in Figure~10, identifying the
top of the southern leg of the loop, corresponds to
the white line in the COR2-B image.  This line naturally passes over
the top of the southern leg of the loop in that image.  Marking
that location in the COR2-B image then identifies the 3-D coordinates
of point \#5.

     The goal is to laboriously do this triangulation for all 15
points for a large set of COR1, COR2, HI1, and HI2 image pairs.  There
are time steps where only some of the 15 points can be properly
triangulated, due to part of the prominence being out of the field of
view in one or both of the STEREO images.  If a given point can be
marked in one of the two images, a 3-D position can be computed
assuming that the trajectory direction of that feature is unchanged
from the most recent time when a proper 2-point triangulation
measurement is available.  This approach ends up being necessary for
all HI2 measurements, because while the prominence is easily visible
in HI2-A (see Figure~3), it is only just barely visible in HI2-B due
to the prominence simply moving too far away from STEREO-B (see
Figure~2).  If a given point is unavailable in either STEREO-A or
STEREO-B image, a position is inferred by interpolation between
3-D positions where actual measurements are available.  All this is
to make sure 3-D position measurements are available for all 15
points at all considered time steps.

     As a final step the distance, $r$, and direction of
each point are inspected to verify that there is a reasonably smooth,
self-consistent variation, where the direction is defined by a
longitude, $\phi$, and a latitude, $\theta$ in HEE coordinates.  To
smooth point-to-point variations, and to allow for further
interpolation between the sampled time steps, sixth order polynomials
are fitted to plots of $\log r$ vs.\ time, $\phi$ vs.\ $\log r$, and
$\theta$ vs.\ $\log r$.  These fits, which are visually confirmed to
be of excellent quality, provide the final measurements of $r$,
$\phi$, and $\theta$ for all 15 points at all desired time steps.

     The results are displayed in Figures~11 and 12.  Figure~11 displays
the evolving prominence structure by projecting the 3-D structure into
2-D image planes in HEE coordinates, with the
first three panels of the figure focusing on the evolution of the
structure close to the Sun, and the last three panels focused farther
out.  The first 10 points are connected with a solid line, as these
represent the well-defined loop (see Figure~10).  As mentioned
above, the connectivity of points \#11-\#15 is less clear.  Dotted
lines are drawn between \#12-\#15, and point \#11 is connected with
\#13, but whether this represents the actual magnetic field line
connectivity is debatable.  It is clear from Figure~11 that the
material represented by points \#11-\#15 erupts at longitudes farther
from Earth than the primary loop, which has a trajectory about
$32^{\circ}$ west of Earth.  For comparison, the CME flux rope of
the eruption is at $40^{\circ}$ west, so the prominence loop is slightly
offset from the center of the CME trajectory, but still well within its
spatial extent.

     This is shown explicitly in Figure~12, which shows the position
of the prominence relative to the CME, based on the separate 3-D
reconstructions of both.  Close to the Sun, the top of the prominence
is at or near the bottom of the apex of the CME flux rope, as apparent
in the COR2 and HI1 images in Figures 3 and 4.  The prominence seems
to move upwards relative to the CME structure, suggesting that
the prominence is not tightly affixed to the CME's magnetic structure.
By the time the prominence is well into the HI2-A field of view, it is
clearly much closer to the leading edge of the CME front relative to
distance from the Sun than is the case in COR2 and early HI1 (see
Figure~3).  Thus, the only way the CME and prominence could be moving
in concert is if the CME structure was not expanding in a self-similar
fashion.  The CME reconstruction assumed self-similar expansion for
the sake of simplicity, and while this is certainly an approximation
it is not apparent that relaxing that approximation would dramatically
improve the quality of the reconstruction.

     This issue can be explored further kinematically.  Figure~13
plots velocities measured for the 15 prominence points as a function
of distance from Sun-center.  The prominence points exhibit little
systematic variation in velocity.  (Note that the waviness of some of
the velocities is an artifact of the polynomial fitting described
above that smoothed out the position and direction measurements.)  The
top of the prominence, represented by points \#5 and \#6, has a speed
of $500-600$ km~s$^{-1}$, with at most only a very small degree of
deceleration.  This kinematic behavior is significantly different from
that of the CME flux rope, for which there is substantial deceleration
in the HI1 field of view, with the CME leading edge ultimately slowing
down from its peak of 1012 km~s$^{-1}$ to 468 km~s$^{-1}$ early in the
HI2 phase (see Figure~6).  With this fundamental difference in
kinematic behavior, the prominence must move upwards relative to the
CME compared to what would be expected from self-similar expansion of
a combined prominence-plus-CME structure.  \citet{tah15b} provided
similar kinematic evidence for the 2012~August prominence also moving
upwards relative to its CME.

     Points \#9 and \#10 have speeds of about 200 km~s$^{-1}$ near
the bottom of the northern leg of the prominence loop.  These are
the points that start to look like comets in HI1-A (see Figure~3).
Thus, the comet-like behavior is demonstrating that the ambient
wind at those locations must be significantly faster than 200 km~s$^{-1}$.

     As far as the shape of the prominence is concerned, there are
no dramatic changes in morphology as the primary loop of interest
expands outwards, but the top of the loop does appear to rotate
slightly at a distance of aboust 50~R$_{\odot}$.  This is most apparent
in Figure~11d and Figure~12.  The effect is modest enough that it is
within the realm of possibility that it could be due to systematic
errors involved in the triangulation measurements.  There are
clearly uncertainties in tracking the same part of
the prominence from COR1 to HI2, which could be systematic in
nature, leading to systematic errors in the inferred
prominence morphology.

     Nevertheless, if the rotation is assumed to be real, there is an
interesting way to quantify it using a quantity called ``writhe,''
as defined by \citet{mab84}.  This quantity has
been used in the study of helicity in solar structures, both
observational and theoretical \citep{dwl97,mgl98,lt05}.
Writhe for the prominence loop can be defined as
\begin{equation}
Wr=\frac{1}{4\pi} \int_0^L \int_0^L \frac{{\bf \hat{t}}(s) \times
  {\bf \hat{t}}(s^{\prime}) \cdot \left[ {\bf X}(s)-{\bf X}(s^{\prime})\right]}
  {\left| {\bf X}(s)-{\bf X}(s^{\prime})\right|^3} ds~ds^{\prime},
\end{equation}
where {\bf X}(s) is the position vector for points along the loop,
and $\hat{\bf t}$ is a tangent vector along the loop.  Calculating
$Wr$ requires integrating over the loop twice, and because writhe
in this manner is defined over a closed loop, we have to connect the
footpoints to close the prominence loop structure.  The $Wr$ parameter
is computed for the 2011~June~7 prominence loop, for all time steps.

     Figure~14 plots $Wr$ versus distance to the top of the loop.
The writhe is initially near zero while the prominence is close to the
Sun, but it increases in magnitude (with a negative sign) between
30~R$_{\odot}$ and 80~R$_{\odot}$.  The writhe reaches a value of
$Wr\approx -0.4$.  The negative magnitude indicates a lefthanded
direction to the rotation.  We attach no meaning to the smaller scale
variations in $Wr$, such as the small bumps near 7~R$_{\odot}$ and
150~R$_{\odot}$, which are too small to be connected with any visible
change in prominence structure in Figures 11 and 12.  But the decrease
in $Wr$ to $\sim -0.4$ clearly corresponds to the rotation of the top
of the loop in Figure~11d and Figure~12.  If real, the cause of this
writhe is unclear.  Prominence eruptions close to the Sun are often
observed to be accompanied by substantial twisting and kinking, but
magnetic forces will be much weaker at distances of
$30-80$~R$_{\odot}$.  The degree of writhe is small enough that it is
possible that some independence of motion of different parts of the
prominence may be enough to induce the apparent rotational motion.

     Shifting focus from the 2011~June event to that of 2012~August,
ideally a full reconstruction analysis would now be presented for the
2012~August eruption.  Unfortunately, this is not possible.  There are
a couple reasons for this.  One is that much of the 2012~August
prominence loop seen in COR1 and COR2 becomes indistinct in HI1 and
HI2, unlike the 2011~June square-topped loop, making it less
attractive for analysis.  But more importantly, the viewing geometry
of the 2012~August event is simply not very good (see Figure~2).  A
point-by-point reconstruction using triangulation requires a clear
association between structures seen in STEREO-A images with structures
seen in STEREO-B images.  This is possible for the 2011~June event,
where both STEREO-A and STEREO-B essentially have lateral views of the
eruption.  But for the 2012~August event STEREO-B has a lateral view
while STEREO-A has very much a backside view (see Figure~2).  This
makes it very difficult if not impossible to identify parts of the
prominence in STEREO-B images with anything in the STEREO-A images.

     The LASCO coronagraphs are actually more helpful for triangulation
purposes, in concert with STEREO-B.  Considering COR2-B and
LASCO/C3 images taken as close in time as possible,
a simple triangulation is made for the top of the apparent prominence
when it is still fully visible, which yields an estimated
longitude of $53^{\circ}$ east of Earth.  This is close to the
$46^{\circ}$ east longitude trajectory inferred for the center of the
flux rope component of the CME (see Figure~2), and also well within the
$42^{\circ}-57^{\circ}$ prominence extent quoted by \citet{tah15b}.
\citet{tah15b} have already performed a kinematic analysis of the
part of the 2012~August prominence structure that is trackable to
1~AU, so kinematic measurements for this event are not presented here.

\section{QUANTIFYING THE SELF-SIMILARITY OF EXPANSION}

     We now use our point-by-point reconstruction of the 2011 June
prominence to assess the degree to which it expands in a self-similar
expansion.  Self-similar expansion is a common assumption for CMEs moving
through the IPM, as it was for us in the CME reconstructions in Section~3.
This assumption greatly
simplifies that type of analysis, as it means that a structure is
morphologically invariant, and all that changes with time is its size.
Nevertheless, it is clear that self-similar expansion is not strictly
true.  Quantifying the degree of deviation from self-similarity is
difficult for CMEs as a whole, because it is not easy to
independently track different parts of CMEs.  Unlike the prominence
material that was the focus of the last section, the visible
structures of CMEs tend to be broad and diffuse, making it difficult
to be sure that a part of the CME seen in an image from one
perspective corresponds to exactly the same part of the CME seen from
a different perspective.  There have nevertheless been attempts to
measure velocities of different parts of CME structures in coronagraph
images to assess the degree of self-similarity at early times.
\citet{bcl87} report a CME velocity field very inconsistent
with self-similar expansion for a fast event from 1980~September~1,
but in a survey of 18 CMEs \citet{dm09} find that
self-similar expansion is usually a good approximation.

     For the erupting 2011~June~7 prominence that was reconstructed
in the last section, we can quantify the degree of
self-similarity more precisely than is possible for CME structures in
general.  Our analysis provides 3-D velocity vectors, ${\bf V}=
(v_r,v_{\theta},v_{\phi})$, at 15 points in the prominence at all
time steps.  Rather than use HEE coordinates as in Figure~11, it is
advantageous to rotate into a coordinate system with the x-axis
pointed towards the center of the prominence, meaning the prominence
is basically propagating in the x-direction.  The average HEE
longitude and latitude of our 15 prominence points over time are
$\phi=36.5^{\circ}$ and $\theta=-2.1^{\circ}$, respectively, so that
is where the x-axis is pointed.  The north ecliptic pole remains in the
xz-plane, meaning the new z-axis will still point in a poloidal
direction, and the y-axis in an azimuthal direction.

     Figure~15 provides a schematic picture of an ovoid structure
expanding away from the Sun.  For such a structure to maintain the
same shape relative to the Sun, velocities within the structure in the
direction of propagation must be proportional to distance from the
Sun.  Thus, in Figure~16a $v_r$ is divided by distance and plotted
versus distance for the 15 measured points at 8 different time steps.
Linear fits are performed to each sequence, with the slope of the
resulting lines indicating deviation from self-similar expansion.
Positive slopes are observed close to the Sun, indicating expansion
greater than that expected for self-similar expansion.  This slope
declines with time, though, and even becomes slightly negative as the
prominence approaches 1~AU.

     Figure~16a provides some sense of the degree of self-similarity,
but it is still desirable to assess self-similarity in a more
quantitative fashion.  A mathematical framework for this is now
provided, with guidance provided by Figure~15.
The object in the figure is shown as an oval, but our prescriptions
will ultimately be independent of shape.  The object is assumed to
have radial and lateral extents defined by the characteristic
distances $a_r$ and $a_{\theta}$ shown in the figure.  We focus first
on the case of pure radial expansion, ignoring lateral expansion for
now.  The front and back sides of the structure are at distances of
$r_1=r+a_r$ and $r_2=r-a_r$.  This object is moving away from the Sun,
with the center, front, and back moving at velocities
$v_r$, $v_{r1}$, and $v_{r2}$, respectively.  If $v_{r1}\neq v_{r2}$
the structure will be changing in size, so $a_r=a_r(r)$.  If it
is assumed to expand (or contract) uniformly relative to its center,
then $v_r=(v_{r1}+v_{r2})/2$.  An aspect ratio for the structure is
defined to be $\Lambda_r(r)=a_r(r)/r$, with differential
\begin{equation}
d\Lambda_r = \frac{1}{r}da_r - \frac{\Lambda_r}{r}dr.
\end{equation}
For self-similar expansion $\Lambda_r$ must be constant ($d\Lambda_r=0$).

     A useful quantification of the self-similarity of the radial
expansion is
\begin{equation}
S_r\equiv \frac{1}{\Lambda_r} \frac{da_r}{dr}=
  \frac{r}{\Lambda_r}\frac{d\Lambda_r}{dr}+1.
\end{equation}
For self-similar expansion, $S_r=1$.  The object will be expanding
faster than self-similar expansion if $S_r>1$, and slower
if $S_r<1$.  A value of $S_r=0$ corresponds to no expansion at all,
and $S_r<0$ means the object will actually be contracting in an
absolute sense as it moves away from the Sun.  Since $v_r=dr/dt$,
\begin{equation}
S_r=\frac{1}{\Lambda_rv_r}\frac{da_r}{dt}.
\end{equation}
From inspection of Figure~15, $da_r/dt=(v_{r1}-v_{r2})/2$,
$v_r=(v_{r1}+v_{r2})/2$, $a_r=(r_1-r_2)/2$,
and $r=(r_1+r_2)/2$, leading to
\begin{equation}
S_r=\frac{(r_1+r_2)(v_{r1}-v_{r2})}{(r_1-r_2)(v_{r1}+v_{r2})}.
\end{equation}
This is the equation that will be used to quantify $S_r$ as a function of
time and distance for the 2011~June~7 prominence.

     The end points of the line segments in Figure~16a are used to
provide the $r_1$, $v_{r1}$, $r_2$, and $v_{r2}$ measurements necessary to
compute $S_r$.  (This is done for all time steps, and not just the 8
shown in the figure.)  Alternatively, measurements at maximum and
minimum distances from the Sun could have simply been used for $r_1$,
$v_{r1}$, $r_2$, and $v_{r2}$.  However, the use of Figure~16a for
this purpose has the advantage of allowing all 15 points to play a
role in measuring $S_r$ at each time step, since all the points
constrain the linear fits.  The slopes of these lines indicate
systematic deviations from self-similarity, while the scatter of the
points about the lines may either indicate more random deviations from
self-similarity within the structure, or may simply indicate
measurement uncertainties.

     In any case, the resulting $S_r$ measurements are shown in
Figure~16d.  The prominence has $S_r=1.3$ close to the Sun at
6~R$_{\odot}$, but this decreases to $S_r=1.1$ by the time the
prominence is fully within the HI1 field of view.  As the prominence
approaches 1~AU, it decreases again, falling slightly below the
self-similar value of 1 before the top reaches 1~AU, fully consistent
with the slope changes seen in Figure~16a.  Despite some degree of
expansion different from self-similarity, the self-similar expansion
approximation is actually not too bad of an approximation for this
erupting prominence, with $S_r$ never venturing too far from unity.

     A similar calculation can be used to assess expansion in the
poloidal direction.  Referring to Figure~15, the aspect ratio is now
$\Lambda_{\theta}(\rho)=a_{\theta}(\rho)/\rho$, where for purposes of
clarity in the figure $\rho$ is now being used as the radial variable
instead of $r$.  The expansion factor is
\begin{equation}
S_{\theta}\equiv \frac{1}{\Lambda_{\theta}} \frac{da_{\theta}}{d\rho}.
\end{equation}
The analog of equation (4) is
\begin{equation}
S_{\theta}=\frac{1}{\Lambda_{\theta}v_{\rho}}\frac{da_{\theta}}{dt}.
\end{equation}
It is important to note that $da_{\theta}/dt$ is affected
by both $v_{\theta}$ and $v_{\rho}$, such that
\begin{equation}
\frac{da_{\theta}}{dt}=v_{\theta}+\Theta v_{\rho},
\end{equation}
where $\Theta$ is the angular half width of the structure.  Noting
also that $\Theta=a_{\theta}/\rho=\Lambda_{\theta}$ (see Figure~15),
this leads to
\begin{equation}
S_{\theta}=\frac{v_{\theta}}{\Theta v_{\rho}}+1.
\end{equation}

     In Figure~16b, $v_{\theta}/v_{\rho}$ is plotted versus the poloidal
angle.  This is done for the same 8 time steps as in Figure~16a,
with linear fits once again made to each set of points.  These lines
should have slopes of zero in the case of self-similar expansion.
The end points of the linear fits in Figure~16b provide
measurements of $v_{\theta1}/v_{\rho1}$, $v_{\theta2}/v_{\rho2}$, $\theta_1$,
and $\theta_2$.  The angular half-width is $\Theta=(\theta_1-\theta_2)/2$,
which is defined relative to a center line that splits the structure into
halves of equal angular width, as in Figure~15.  For the upper half,
we have from equation (9)
\begin{equation}
S_{\theta1}=\frac{v_{\theta1}}{\Theta v_{\rho1}}+1.
\end{equation}
For the lower half, we have
\begin{equation}
S_{\theta2}=-\frac{v_{\theta2}}{\Theta v_{\rho2}}+1,
\end{equation}
where the negative sign comes from recognizing that for the lower half
of the structure a positive $v_{\theta}$ corresponds to movement towards
the center line (i.e., a contraction) instead of movement away from
the center line (i.e., an expansion).  Our estimate of $S_{\theta}$ for
the whole structure is simply the average of $S_{\theta1}$ and $S_{\theta2}$,
\begin{equation}
S_{\theta}=\frac{1}{(\theta_1-\theta_2)}\left(\frac{v_{\theta1}}{v_{\rho1}}-
  \frac{v_{\theta2}}{v_{\rho2}}\right)+1.
\end{equation}
providing the $S_{\theta}$ curve in Figure~16d.
Likewise, for the azimuthal direction, an expansion
factor $S_{\phi}$ can be computed,
\begin{equation}
S_{\phi}=\frac{1}{(\phi_1-\phi_2)}\left(\frac{v_{\phi1}}{v_{\rho1}}-
  \frac{v_{\phi2}}{v_{\rho2}}\right)+1,
\end{equation}
leading to the $S_{\phi}$ curve in Figure~16d.

     As is the case for $S_r$, $S_{\theta}$ and $S_{\phi}$ remain relatively
close to the self-similar value of 1, once again consistent with the
notion that self-similar expansion is a decent approximation.  Both
$S_{\theta}$ and $S_{\phi}$ exhibit more variability than $S_r$, but
it is possible that this is simply due to measurement uncertainty.
Uncertainties in $S_{\theta}$ and $S_{\phi}$ will be larger than those
in $S_r$, because while there is plenty of radial extent to the
prominence, along which one can measure the radial velocity gradient,
there is not much angular extent (see Figure~11), making it harder to
accurately quantify the lateral velocity gradients across the width of
the structure.

     All three expansion factors are near 1.2 at distances of
about 10~R$_{\odot}$, implying expansion somewhat greater than
self-similarity while the prominence is in the COR2 field of view.
The largest deviation from self-similarity is for
$S_{\theta}$ far from the Sun, which plunges below $S_{\theta}=0.6$ as it
approaches 1~AU.  This is also indicated by the negative slopes of
the lines in Figure~16b that represent the last time steps shown.
This suggests expansion perpendicular to the ecliptic plane is
being suppressed to some extent far from the Sun.

\section{THE SCARCITY OF PROMINENCE MATERIAL AT 1 AU}

     The bright and compact appearance of the prominence structures
in HI2 is quite distinctive from the broader, more diffuse emission
generally observed in fronts of CMEs and CIRs.  This means that it is
possible to systematically search HI2 images for other examples of
prominence material out near 1~AU.  A search of the entire 2007--2014
STEREO database of HI2 images was made for this purpose.  However, no
other examples were found.  It should be emphasized that while the
database search should certainly have detected cases as bright and
obvious as the 2011 June and 2012 August eruptions, it could
have missed fainter examples.  Nevertheless, the failure to find other
clear examples of 1~AU prominence eruptions emphasizes that this is
not a common phenomenon.  This conclusion is consistent with in~situ
studies that find that it is rare to detect unambiguous cool
prominence material within CMEs.

     Looking within in~situ CME encounters with ACE, \citet{stl10}
identified regions of low charge state and
low temperature with possible prominence material, finding that
only 4\% of CMEs encountered are observed to have such material.
This makes prominence material in CMEs at 1~AU already seem rather
rare, but events analogous to the two discussed in this article
will be even more unusual.  The 4\% of cases found by \citet{stl10}
are not generally accompanied by large density
enhancements, including the published example, which shows only a
very weak density increase.  There can be
no question that the 2011~June~7 and 2012~August~31 cases would
have produced far more dramatic density increases than this if
the prominences had encountered a spacecraft at 1~AU.

     One would be hard pressed to find a published example of what
an in~situ encounter with the 1~AU events discussed here might
look like.  The 1997~January~10 event observed by {\em Wind} and
discussed by \citet{lb98} may be one example.  This event
shows an extremely large density enhancement at the back end of a
magnetic cloud ($n_p\approx 150$ cm$^{-3}$), with a low temperature.
\citet{lb98} associate this with prominence material collected at the
back of the CME flux rope, consistent with what one might expect if the
2011~June~7 prominence had been encountered by {\em Wind}.  One
problem with this interpretation is that the 1997~January~10 CME
is followed by a CIR structure, meaning that the density enhancement
may also be a pile-up associated with the CIR.

\section{SUMMARY}

     Observations from the STEREO mission have been used to
study prominence eruptions that can be tracked continuously
from the Sun all the way to 1~AU.  The results of this study
can be summarized as follows.
\begin{description}
\item[1.] A full search of the HI2 image database, covering the
  years 2007-2014, found only two examples of clear, bright, 
  and compact prominence emission close to 1~AU.  These are
  eruptions from 2011~June~7 and 2012~August~31.
  The lack of many cases of prominence eruptions being
  trackable all the way to 1~AU is consistent with the rarity
  of cases of in~situ instruments at 1~AU encountering
  cold, dense prominence material.
\item[2.] Comprehensive STEREO observations of the two events and
  their journey through the IPM are here provided.
  A complete loop is apparent for both structures close to
  the Sun, but only for the 2011~June event can the full loop
  be tracked far into the IPM.  For the 2011~June event, knots
  of material far down one leg of the loop begin to look like
  comets in the HI1-A field of view, as the ambient solar wind
  attempts to pick up and accelerate the trailing prominence
  material.
\item[3.] A thorough reconstruction of the expanding prominence
  structure is performed for the 2011~June event, using
  point-by-point triangulation, allowing the morphological
  and kinematic evolution of the structure to be inferred from
  the Sun to 1~AU.  Unfortunately, such an analysis is not possible
  for the 2012~August event due primarily to a poor viewing geometry.
\item[4.] Comprehensive 3-D morphological and kinematic reconstructions
  are performed for the CMEs that accompany the two prominence
  eruptions, modeling the CME ejecta with flux rope shapes.
  Both are fast CMEs, reaching speeds greater than 1000 km~s$^{-1}$,
  and both have visible shocks in front, the shapes of which are
  also modeled in the 3-D reconstructions.  Both
  erupting prominences are directed within $10^{\circ}$ of the
  central trajectory of the CME flux rope.
\item[5.] The CME reconstruction includes constraints from in~situ
  instruments at 1~AU.  The flanks of the 2012~August~31 CME shock
  hit STEREO-B and {\em Wind}.  For the 2011~June~7 CME, {\em Wind}
  observes a density increase that is probably associated with a grazing
  incidence encounter with the CME.  The CME arrival times
  indicated by the in~situ data are well matched by the
  3-D CME reconstructions for both events.  For the 2012~August event,
  shock jump conditions provide measurements
  of the shock normal at STEREO-B and {\em Wind}, which
  agree well with the shock normals predicted by the image-based
  3-D reconstruction of the shock shape.
\item[6.] The 2011~June~7 prominence moves upwards relative to the
  larger CME structure enveloping it, implying that the prominence
  structure is not tightly affixed to the CME's magnetic structure.
  The prominence kinematics are significantly different from
  the CME, with the CME experiencing more deceleration than any part
  of the prominence.
\item[7.] The top of the 2011~June~7 prominence loop appears to rotate
  slightly as it moves through the HI1-A field of view, with
  the writhe reaching $Wr\approx -0.4$.  If not due to systematic
  uncertainties in the reconstruction, this could be 
  due to deformation from different parts of the loop having different
  nonradial velocity components rather than magnetic forces.
\item[8.] Analysis of the expansion of the 2011~June~7 prominence
  structure yields a quantification of its degree of self-similarity,
  in both radial and lateral directions.
  Self-similar expansion appears to be a decent approximation for
  this structure, with the the expansion factors $S_r$, $S_{\theta}$,
  and $S_{\phi}$ remaining between 0.75 and 1.25 for most of the
  prominence's journey to 1~AU.  The radial expansion is
  persistently greater than self-similar until the prominence
  approaches 1~AU.
\end{description}



\acknowledgments

Financial support was provided by the Chief of Naval Research,
and by NASA/LWS award NNH14AX61I to the Naval Research Laboratory.

\clearpage

\begin{figure}[t]
\plotfiddle{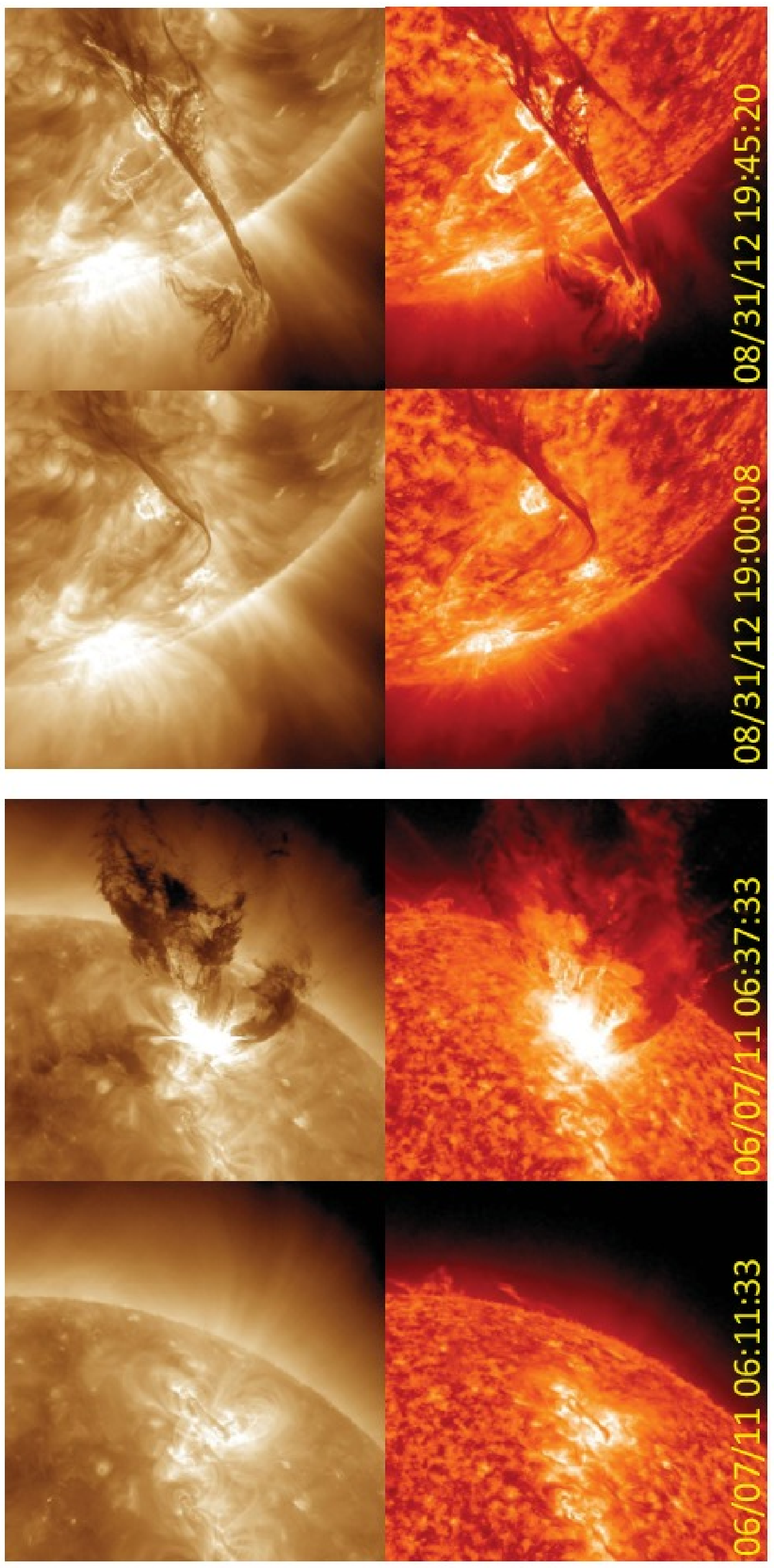}{2.5in}{-90}{65}{65}{-260}{350}
\caption{SDO/AIA images of the prominence eruptions from 2011~June~7 (left)
  and 2012~August~31 (right), showing a single pre- and post-eruption image
  each, for the Fe~XII $\lambda$193 (top) and He~II $\lambda$304 (bottom)
  bandpasses.}
\end{figure}

\clearpage

\begin{figure}[t]
\plotfiddle{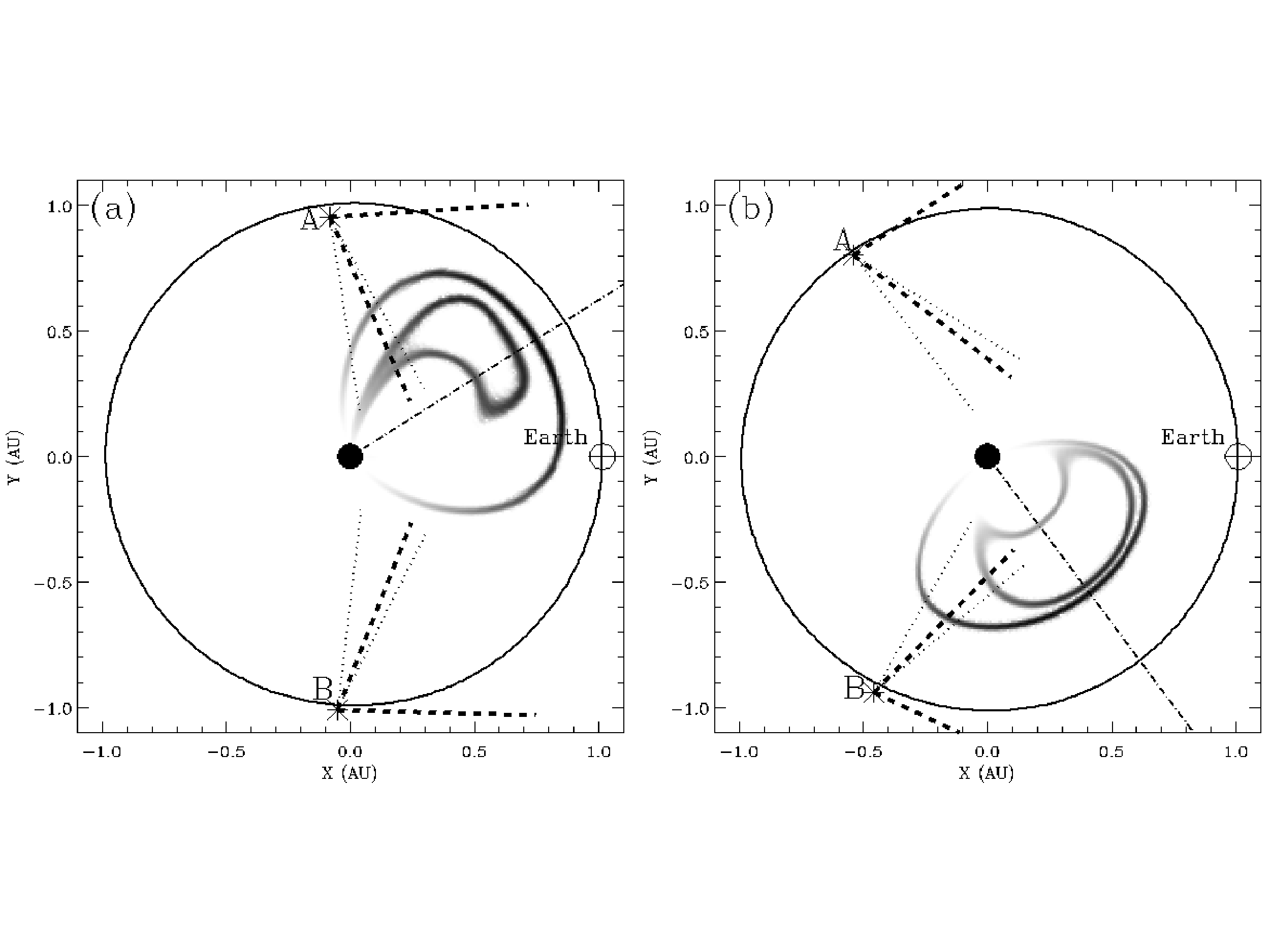}{2.5in}{0}{65}{65}{-235}{-60}
\caption{(a) Ecliptic plane map at the time of the 2011~June~7 CME and
  prominence eruption in heliocentric earth ecliptic coordinates,
  showing the positions of STEREO-A, STEREO-B, and Earth.  The spatial
  extent of the CME in the ecliptic plane is indicated using a slice
  through the 3-D reconstruction described in Section~3.  The
  dot-dashed line indicates the trajectory of the prominence eruption
  (see Section~4).  Dotted and dashed lines indicate the fields of
  view of the HI1 and HI2 heliospheric imagers onboard the STEREO
  spacecraft, respectively. (b) Similar to (a), but for the
  2012~August~31 event.}
\end{figure}

\clearpage

\begin{figure}[t]
\plotfiddle{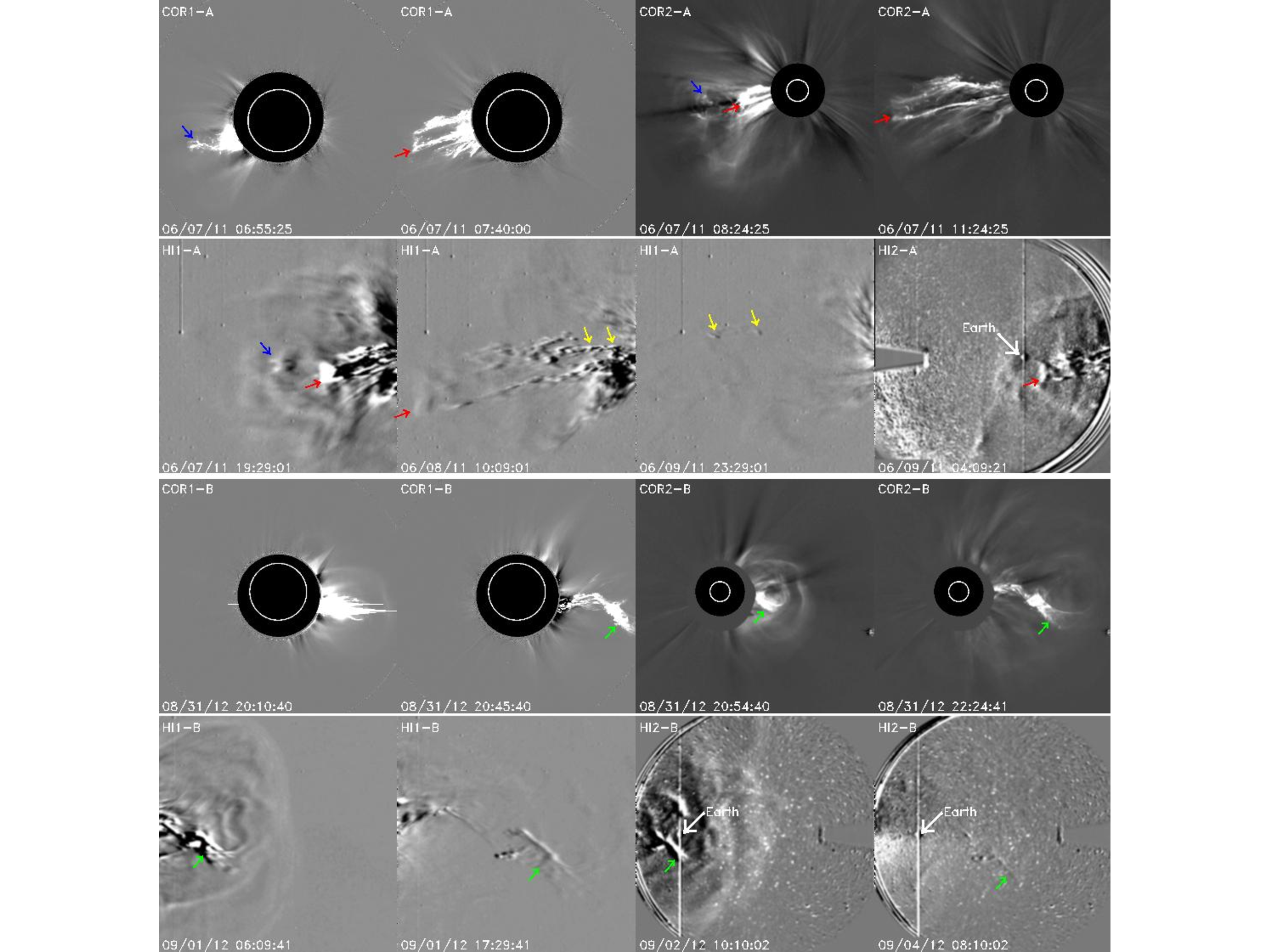}{5.5in}{0}{85}{85}{-305}{0}
\caption{The top two rows are a sequence of 8 images from STEREO-A
  depicting the 2011 June 7 prominence eruption as it journeys from
  the Sun into the IPM.  Similarly, the bottom two rows are a sequence of
  STEREO-B images showing the progression of the 2012 August 31
  prominence eruption.  Colored arrows identify specific parts
  of the prominences that can be tracked from frame to frame.  Note that
  there are saturation artifacts in the earliest COR1 images.
  More comprehensive visualizations of the prominence propagation
  through the fields of view of the STEREO imagers are provided in two
  movies available in the online edition of the article.}
\end{figure}

\clearpage

\begin{figure}[t]
\plotfiddle{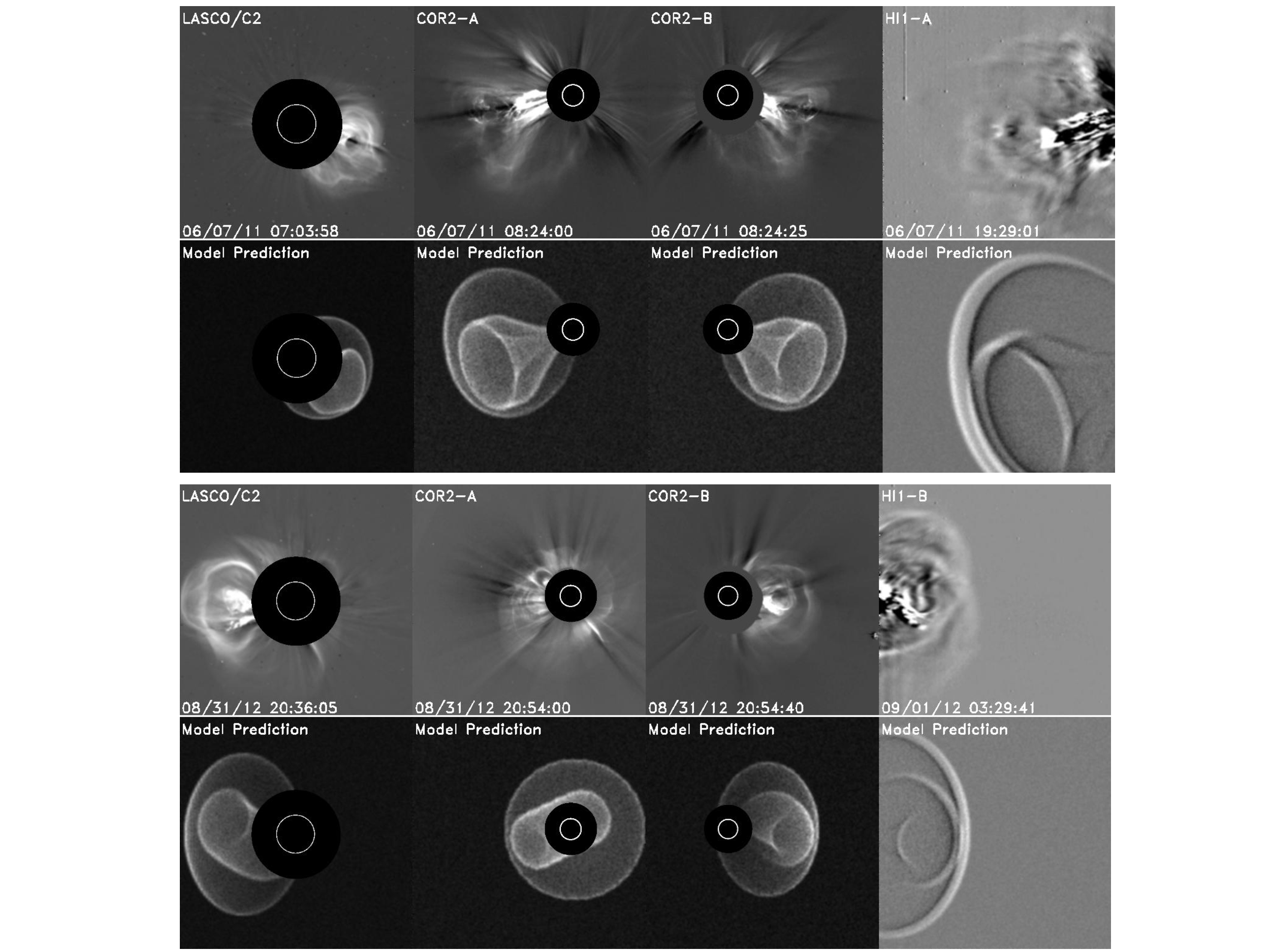}{5.5in}{0}{85}{85}{-310}{0}
\caption{In the top two rows, four images of the 2011 June 7 CME
  are compared with the predictions of the 3-D model from Figure~5.  The
  bottom two rows are an analogous data/model comparison for four
  images of the 2012 August 31 CME.}
\end{figure}

\clearpage

\begin{figure}[t]
\plotfiddle{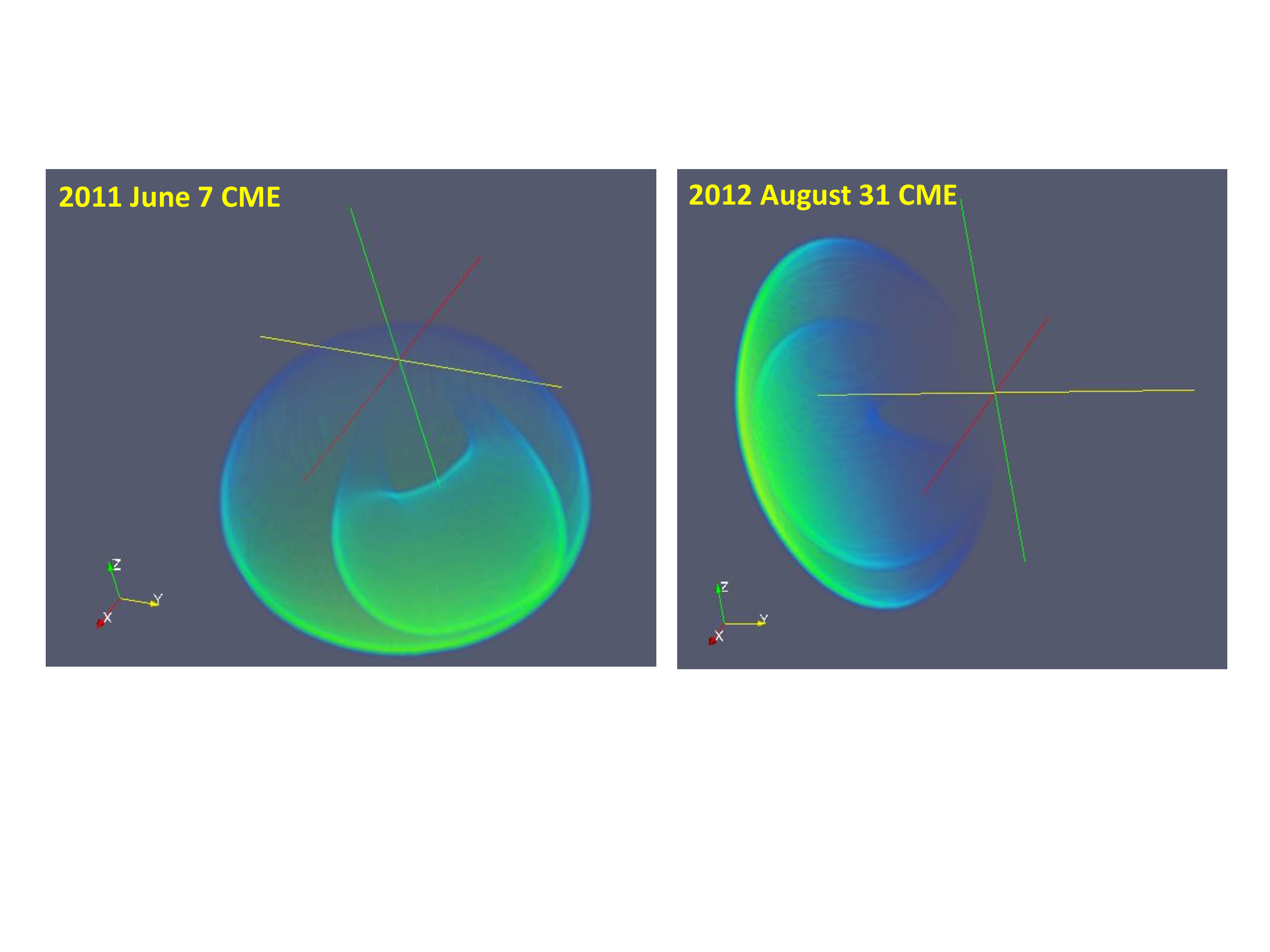}{1.5in}{0}{68}{68}{-245}{-100}
\caption{3-D reconstructions of the 2011 June 7 and 2012 August 31 CME,
  structures in HEE coordinates, with the x-axis in red pointing towards
  Earth, and the z-axis in green pointing towards ecliptic north.  Each
  model consists of a flux rope shape embedded inside a lobular shock.}
\end{figure}

\clearpage

\begin{figure}[t]
\plotfiddle{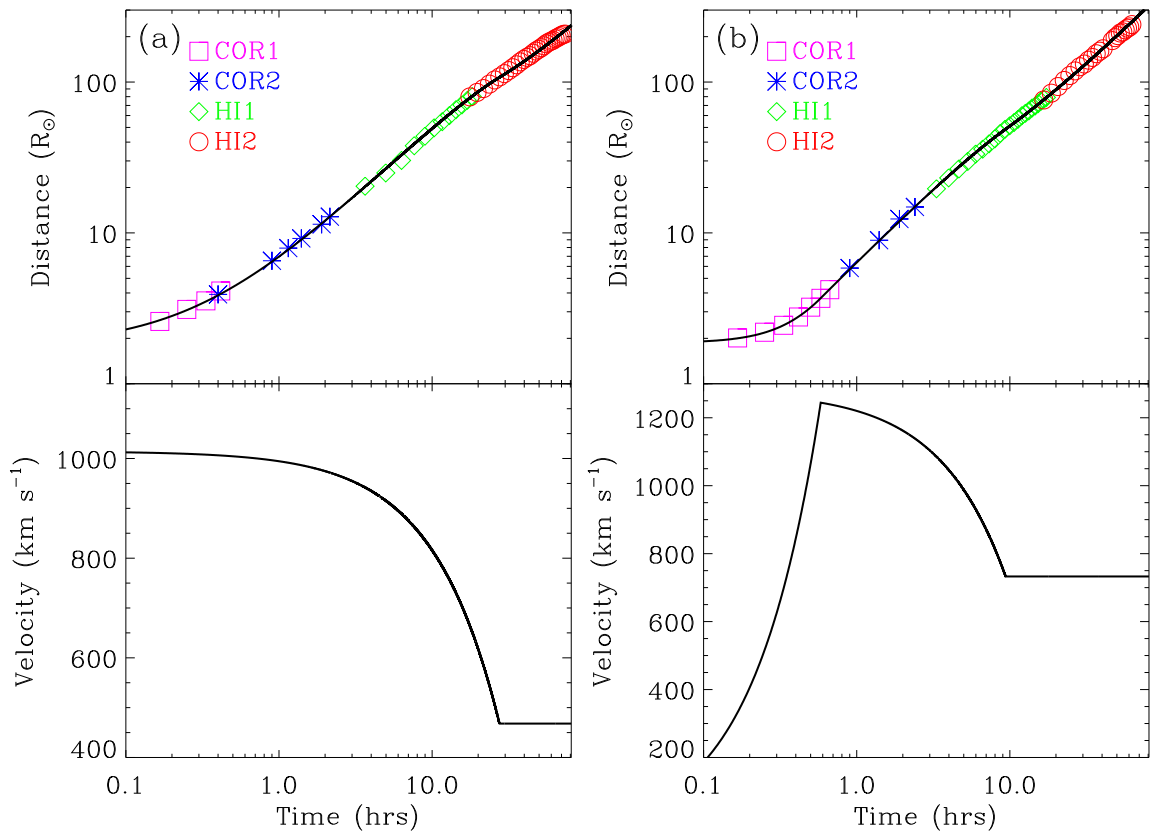}{2.5in}{0}{90}{90}{-270}{-320}
\caption{(a) Kinematic model of the 2011 June 7 CME based on measurements
  from the imagers on STEREO-A.  The model follows the leading edge of the
  flux rope component of the CME.  (b) Kinematic model of the 2012
  August 31 CME based on measurements from STEREO-B, following the leading
  edge of its flux rope component.}
\end{figure}

\clearpage

\begin{figure}[t]
\plotfiddle{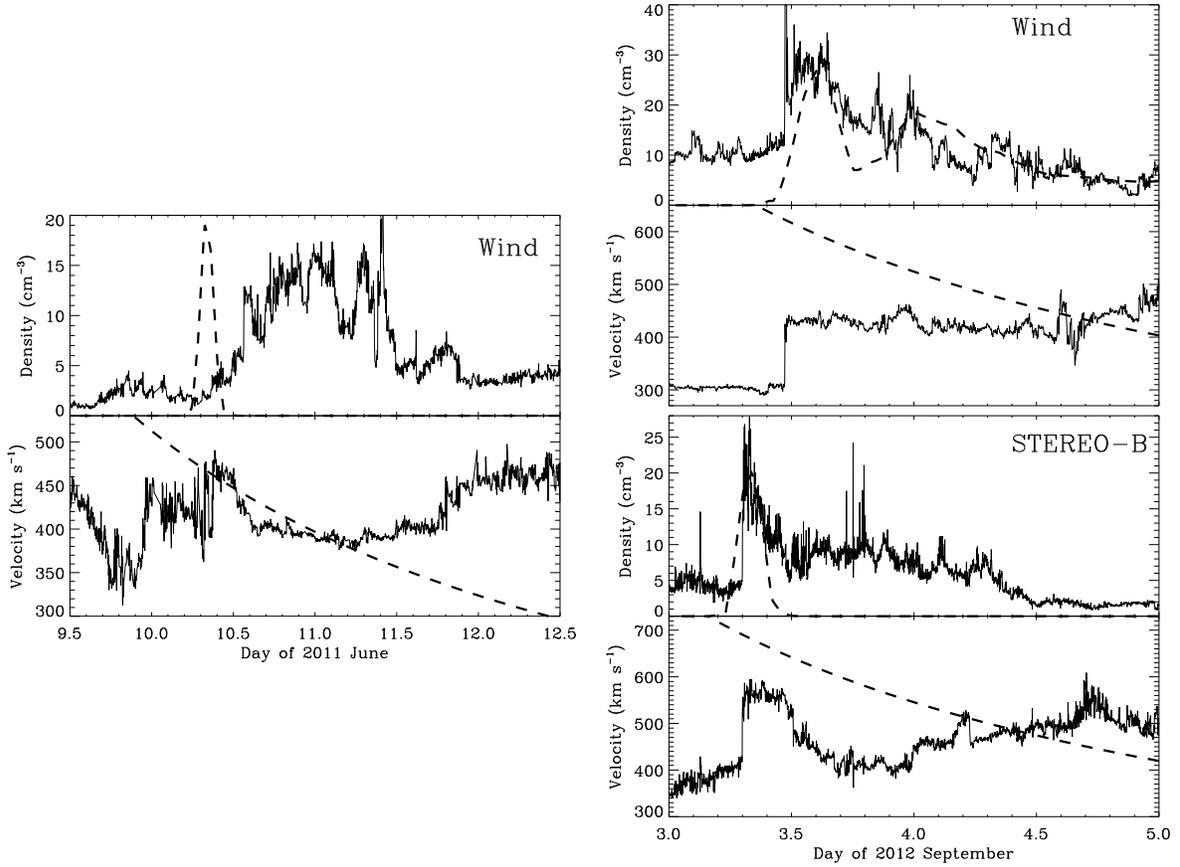}{2.5in}{90}{75}{75}{310}{-60}
\caption{The left panels show in~situ density and velocity measurements
  from {\em Wind} during the time in which the 2011~June~7 CME may have
  encountered the spacecraft.  The right panels show in~situ density and
  velocity measurements from {\em Wind} and STEREO-B, for the time in
  which these spacecraft encountered the 2012~August~31 CME.  The dashed
  lines in all panels are the density and velocity profiles predicted by
  the 3-D reconstructions of both CMEs described in Section~3.}
\end{figure}

\clearpage

\begin{figure}[t]
\plotfiddle{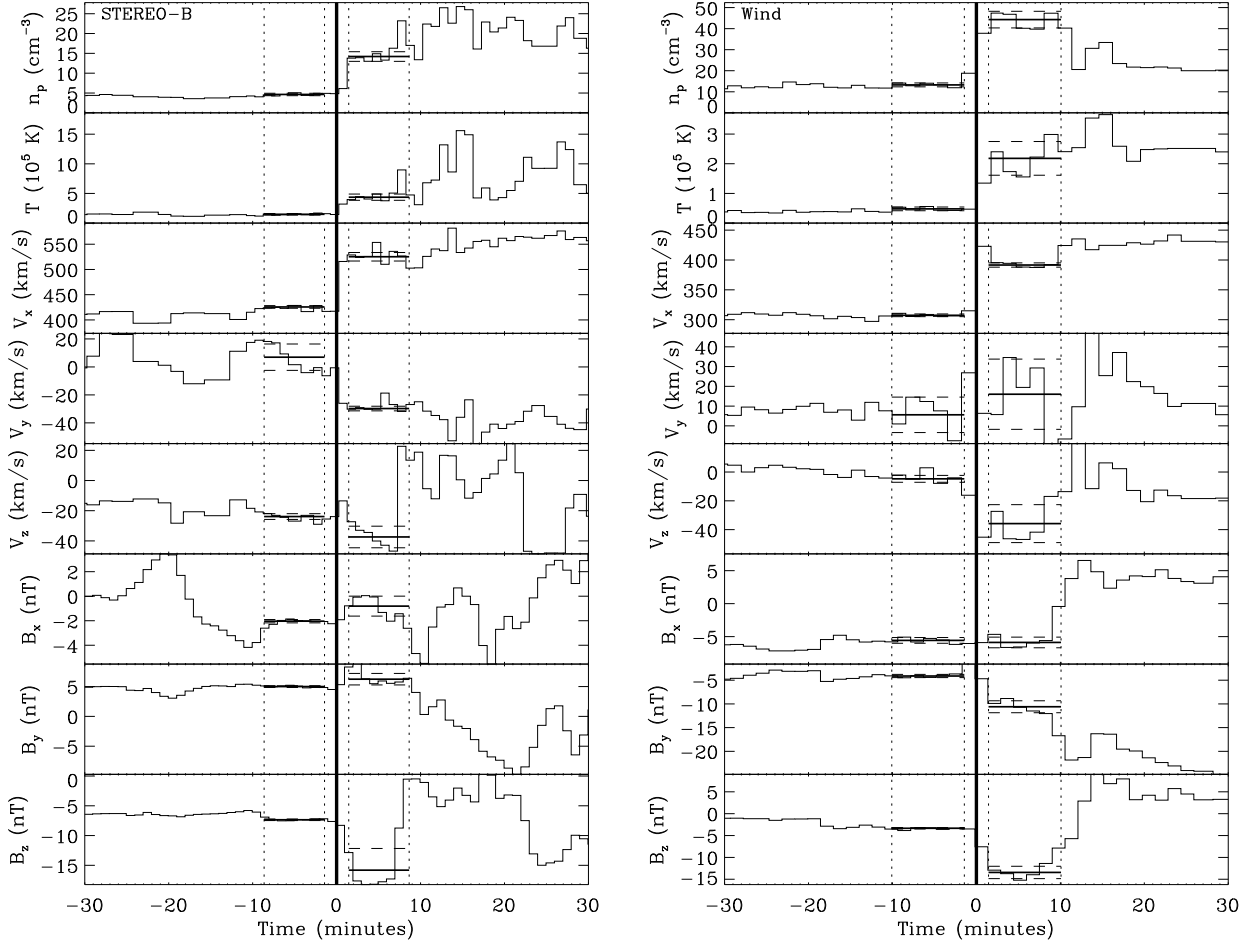}{5.0in}{90}{75}{75}{295}{-50}
\caption{In~situ observations of the 2012~August~31 CME shock at both
  STEREO-B and {\em Wind}.  From top to bottom the quantities are
  proton density, temperature, three components of velocity, and three
  components of magnetic field.  The velocity and field components are
  for a spacecraft-centered RTN coordinate system with the x-axis pointing
  away from the Sun.  Dotted lines indicate time ranges
  used to measure pre- and post-shock plasma parameters.  These
  measurements and their uncertainties are shown as horizontal solid
  and dashed lines, respectively.}
\end{figure}

\clearpage

\begin{figure}[t]
\plotfiddle{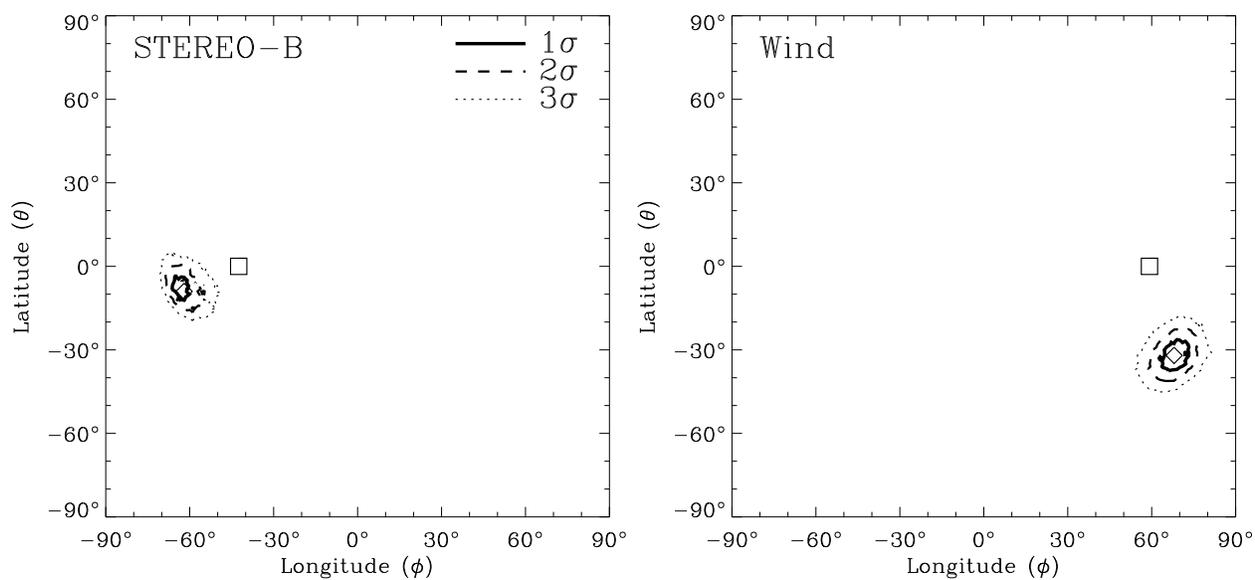}{3.0in}{90}{75}{75}{295}{-100}
\caption{Shock normals relative to the Sun-observer axis, computed for
  the 2012~August~31 CME shock at STEREO-B (left) and {\em Wind}
  (right), based on the plasma measurements in Figure~8 and the
  Rankine-Hugoniot shock jump conditions.  Diamonds indicate the best
  fit, with surrounding 1$\sigma$, 2$\sigma$, and 3$\sigma$ confidence
  intervals.  Squares indicate the expected shock normal based on
  the image-based 3-D reconstruction in Figure~5.}
\end{figure}

\clearpage

\begin{figure}[t]
\plotfiddle{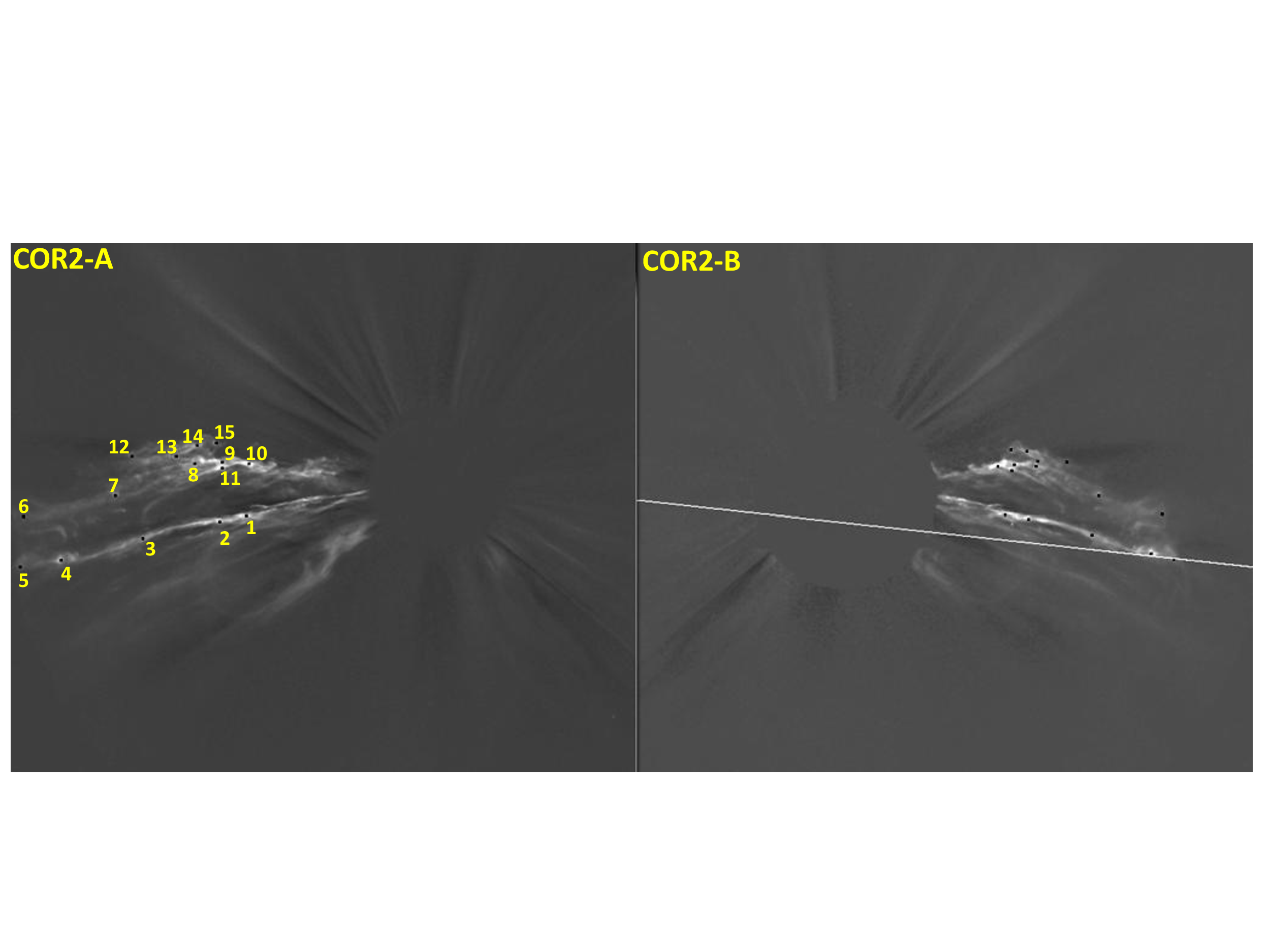}{2.5in}{0}{65}{65}{-233}{-60}
\caption{The left panel shows a COR2-A images of the 2011~June~7 erupting
  prominence and marks 15 points that are tracked in STEREO imagery
  as the prominence expands away from the Sun.  The right panel is a COR2-B
  image taken at the same time as the COR2-A image, with the same 15
  points identified (but not labeled).  The line of sight corresponding to
  point \#5 in COR2-A, the top of the southern leg of the structure,
  corresponds to the white line in the COR2-B panel, which also crosses
  the top of the southern leg.}

\end{figure}

\clearpage

\begin{figure}[t]
\plotfiddle{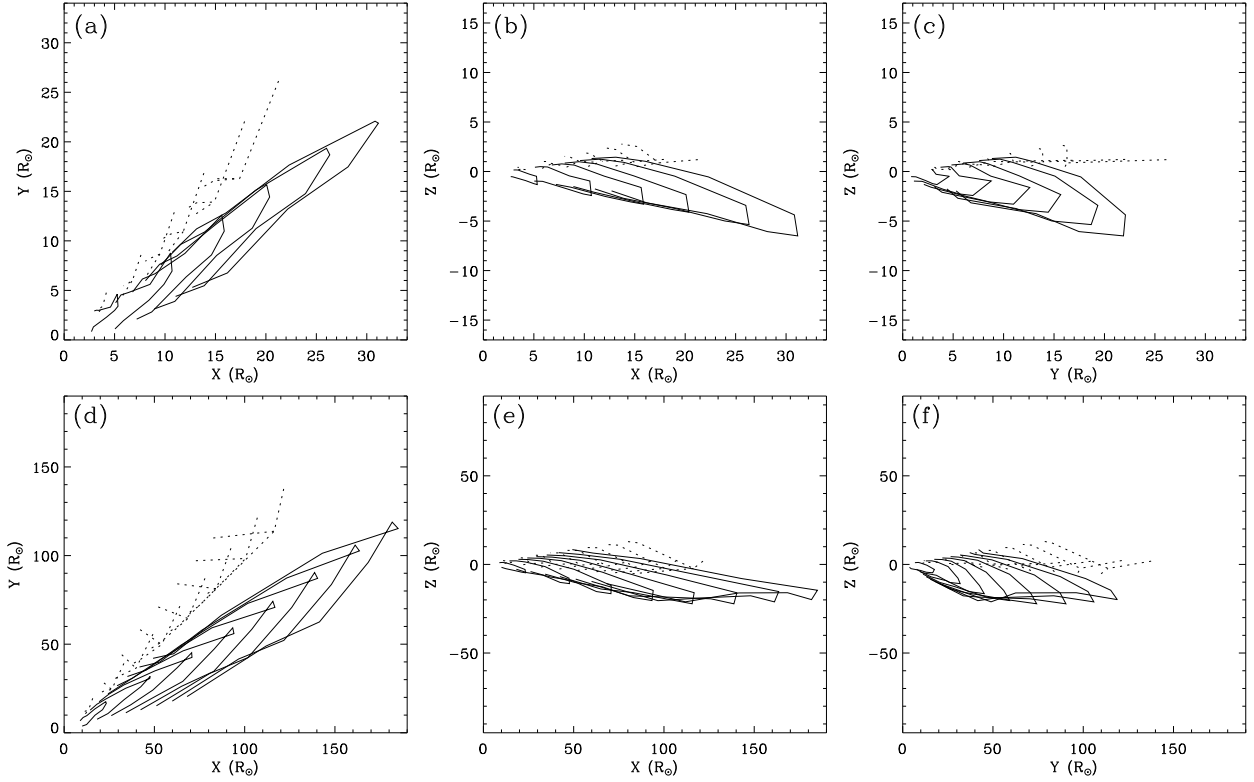}{2.5in}{90}{70}{70}{275}{-70}
\caption{Evolution of the 2011~June~7 prominence eruption, from a 3-D
  reconstruction based on point-by-point triangulation, with panels
  (a)-(c) focusing on evolution close to the Sun and panels (d)-(f)
  focused further out in the IPM.  The solid lines trace a loop
  structure seen in the eruption, based on points \#1-\#10 in
  Figure~10, while the dotted lines are based on points \#11-\#15 in
  Figure~10, with an assumed connectivity between them that is more
  questionable.  The maps are in HEE coordinates,
  with the x-axis pointed towards Earth and the z-axis towards the
  north ecliptic pole.  In panels (a)-(c) the 6 maps correspond to
  times of $\Delta t$=2.15, 4.33, 6.50, 8.40, 10.90, and 12.90 hours
  from a reference time of 6:30 UT on 2011~June~7.  In panels (d)-(f),
  the 8 maps correspond to times of $\Delta t$=9.66, 19.65, 28.98,
  38.32, 47.66, 57.66, 67.65, and 76.98 hours.}
\end{figure}

\clearpage

\begin{figure}[t]
\plotfiddle{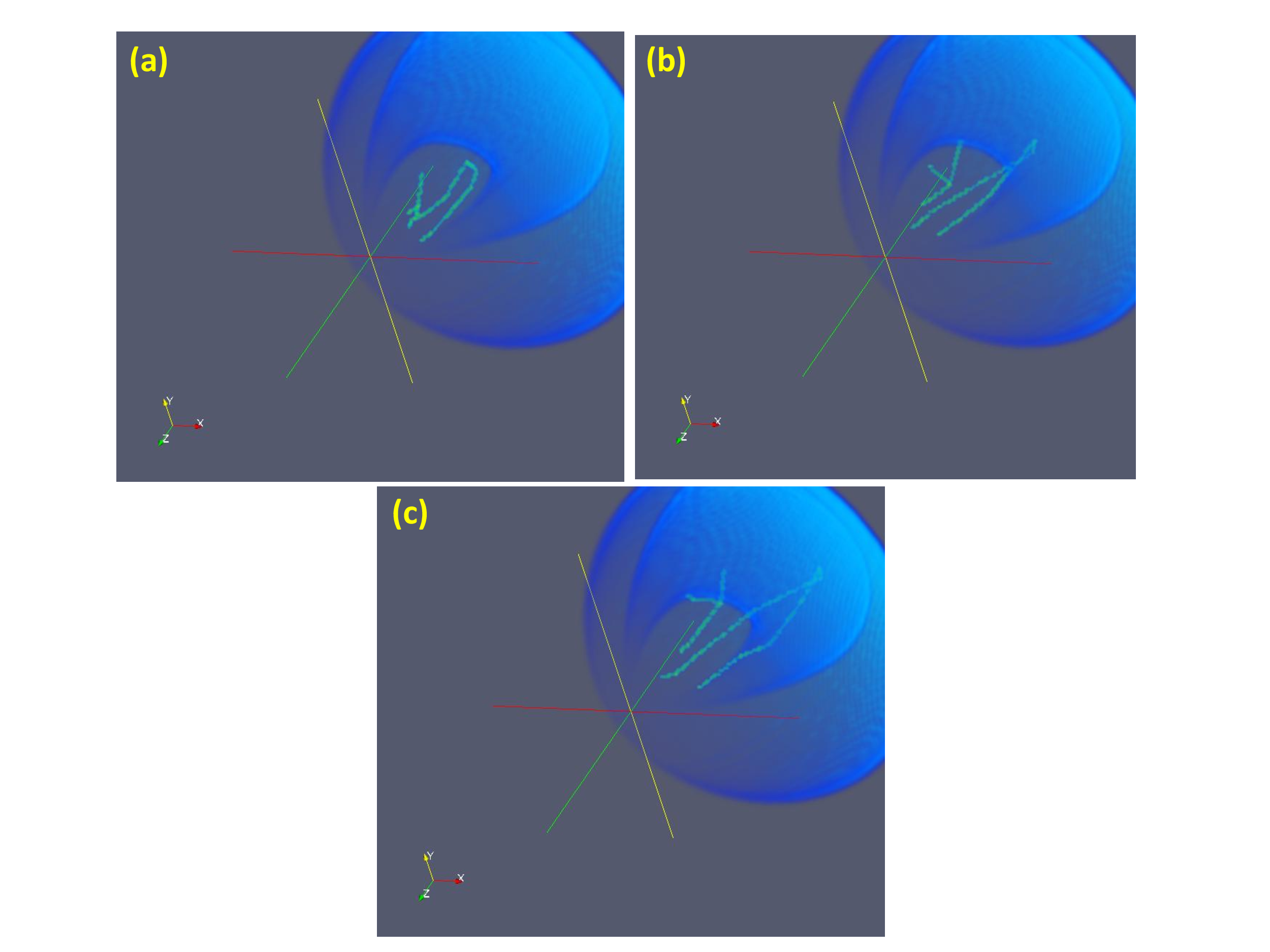}{4.5in}{0}{77}{77}{-275}{-15}
\caption{(a) For the 2011~June~7 eruption, the position of the evolving
  prominence structure is shown relative to the reconstructed CME
  structure, for a time corresponding to $\Delta t$=5.40 hours from
  a reference time of 6:30 UT on 2011~June~7.  The coordinate system is
  HEE, with the x-axis in red pointing towards Earth, and the z-axis in
  green pointing towards ecliptic north.  (b) Similar to (a) but for
  $\Delta t$=27.66 hours.  (c) Similar to (a) but for $\Delta t$=56.66
  hours.  The top of the prominence appears to rotate slightly,
  and the prominence moves upwards with time relative to the CME.}
\end{figure}

\clearpage

\begin{figure}[t]
\plotfiddle{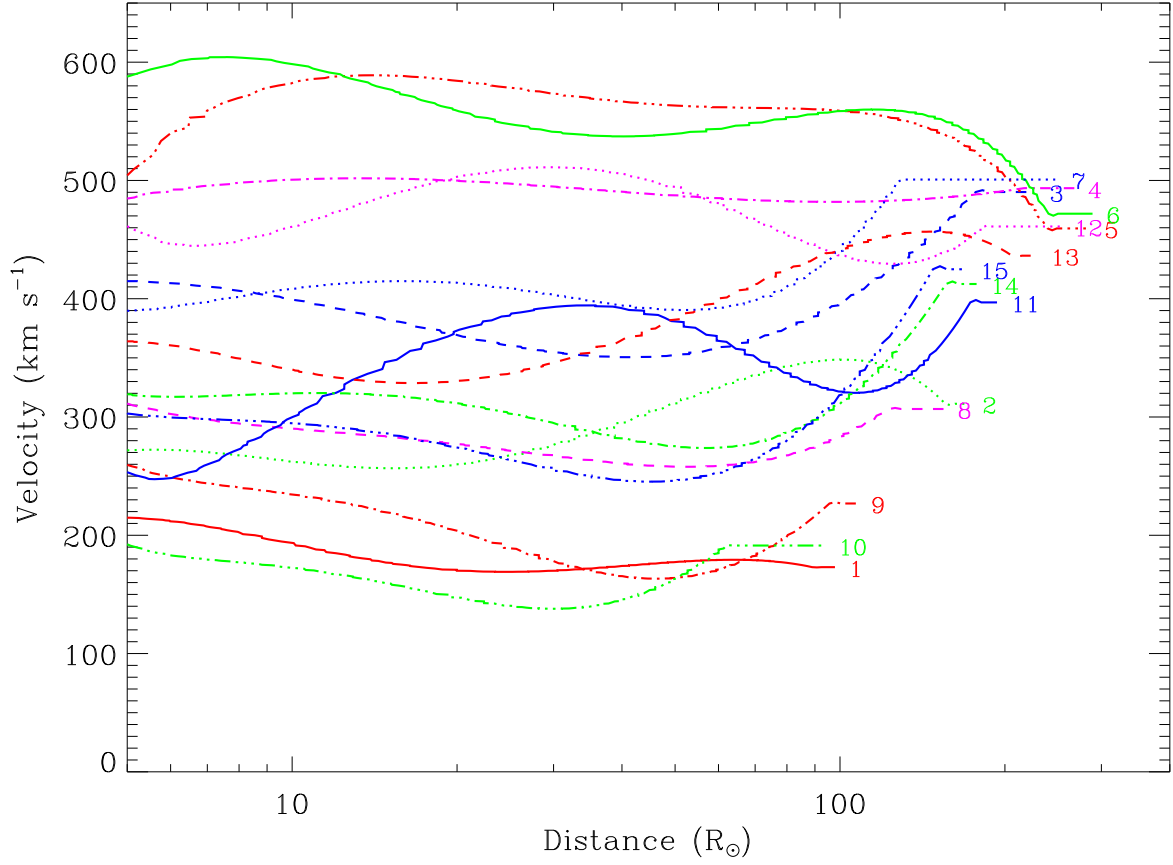}{2.0in}{90}{70}{70}{275}{-50}
\caption{The speeds of the 15 points followed in the 2011~June~7
  prominence (see Figure~10) are plotted versus distance from Sun-center.}
\end{figure}

\clearpage

\begin{figure}[t]
\plotfiddle{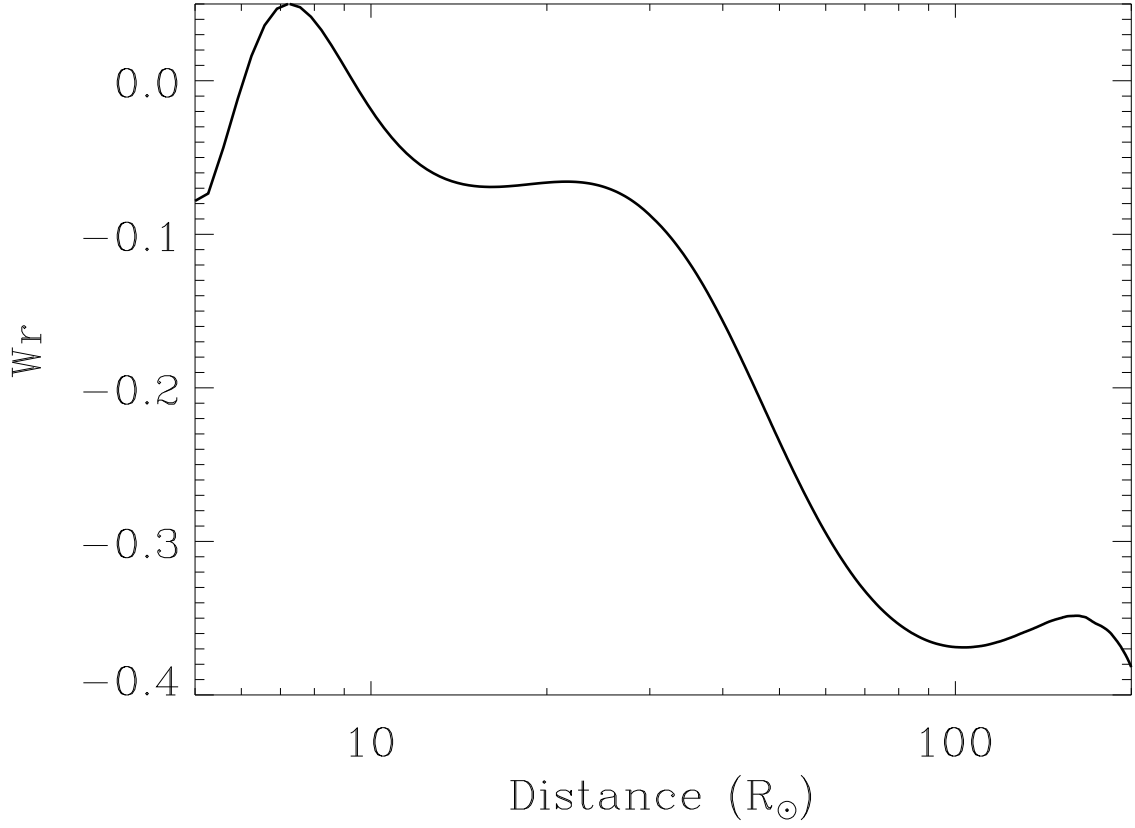}{2.0in}{90}{70}{70}{275}{-50}
\caption{The writhe of the 2011~June~7 prominence loop plotted as
  a function of Sun-center distance to the top of the loop.}
\end{figure}

\clearpage

\begin{figure}[t]
\plotfiddle{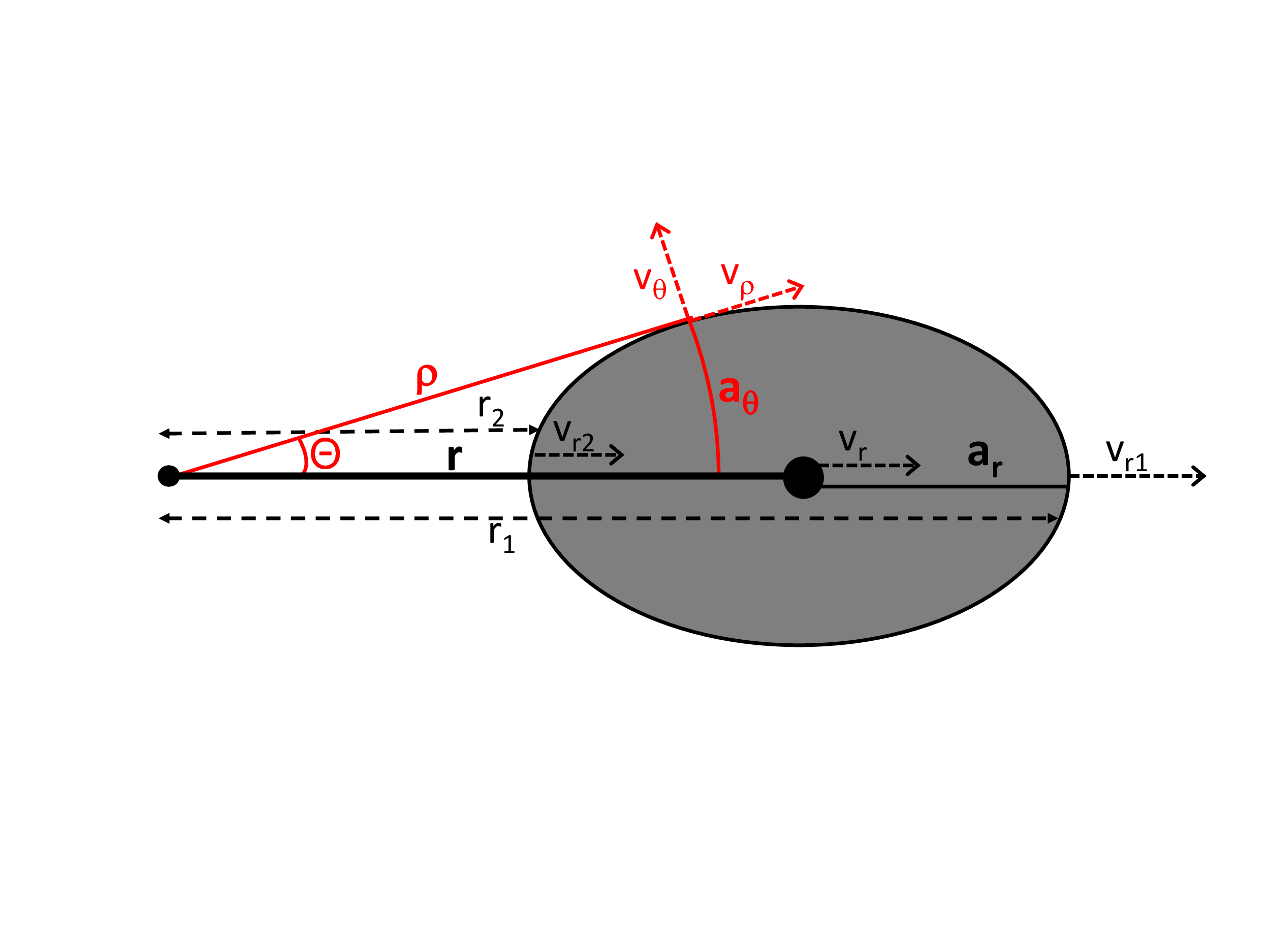}{1.0in}{0}{67}{67}{-260}{-60}
\caption{Schematic picture of an expanding structure at a distance $r$
  from the Sun.}
\end{figure}

\clearpage

\begin{figure}[t]
\plotfiddle{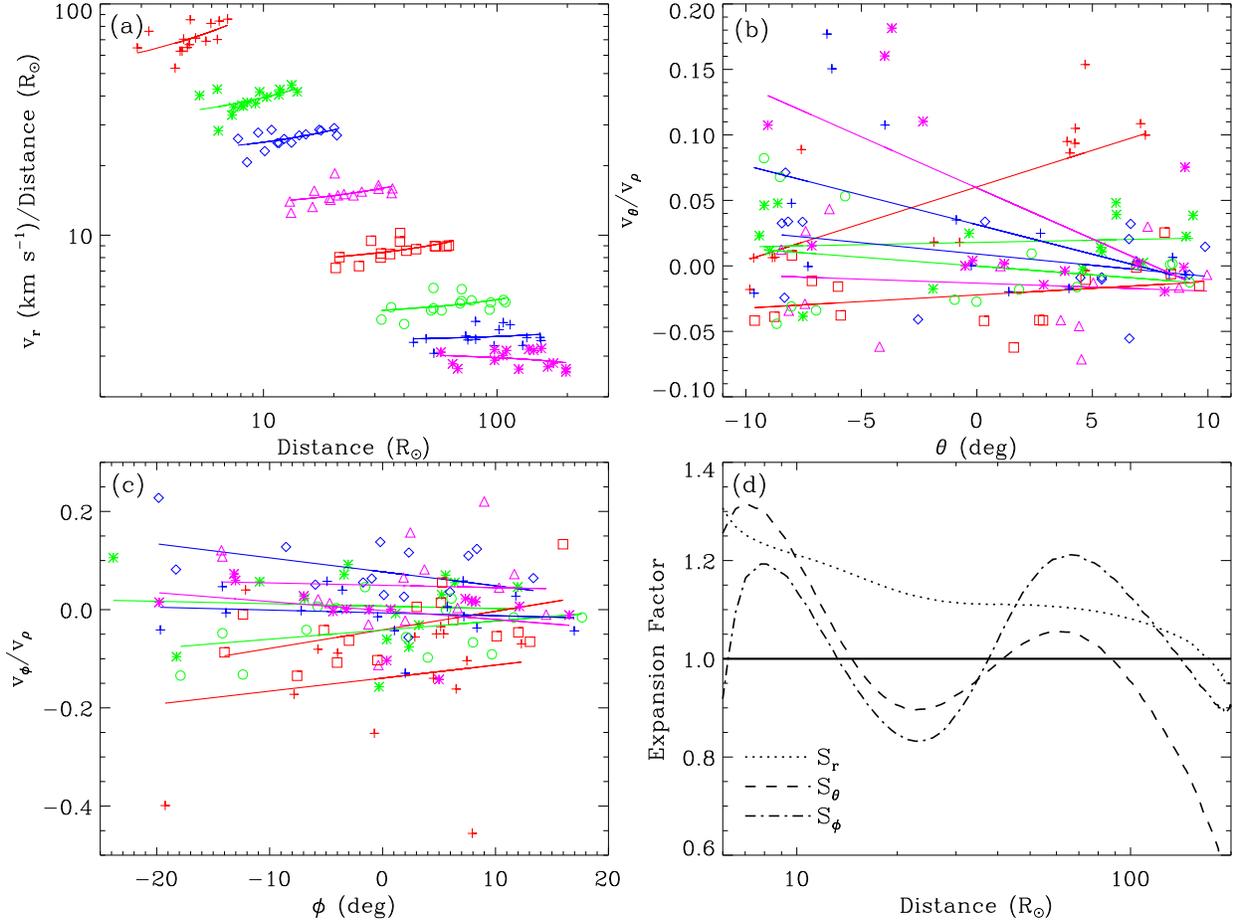}{4.0in}{90}{70}{70}{275}{-40}
\caption{(a) For the 15 points followed in the 2011~June~7 prominence
  (see Figure~10), the radial velocity ($v_r$) divided by the distance
  is plotted versus Sun-center distance.  This is done for 8 different
  time steps, and a linear fit is performed to each 15-point sequence.
  The slope of these lines is indicative of the degree of
  self-similarity of the expanding structure, with a slope of 0
  corresponding to self-similar radial expansion.  (b) For the same 8
  time steps, the poloidal velocity divided by the radial velocity
  ($v_{\theta}/v_{\rho}$) is plotted versus
  poloidal angle $\theta$, and lines are again fitted to each 15-point
  sequence, with zero slope corresponding to self-similar expansion.
  Line colors and symbols are kept
  consistent with those used in (a).  (c) Analogous to (b), but for
  the azimuthal velocity ($v_{\phi}$).  (d) The expansion factors
  in the radial, azimuthal, and poloidal directions ($S_r$,
  $S_{\theta}$, and $S_{\phi}$) are plotted versus distance to the top
  of the prominence structure.  The horizontal line corresponds to
  self-similar expansion.}
\end{figure}

\end{document}